\pdfoutput=1

\documentclass[11pt]{article}

\usepackage{acl}

\usepackage{times}
\usepackage{latexsym}

\usepackage[T1]{fontenc}

\usepackage[utf8]{inputenc}

\usepackage{microtype}

\usepackage{inconsolata}
\usepackage{multirow}
\usepackage{amsmath}
\usepackage{csquotes}
\usepackage{twemojis}

\usepackage{pgfplots,pgfplotstable}
\tikzset{
  font={\fontsize{8pt}{10}\selectfont}}
\usepackage{graphicx}
\usepackage{subfigure}

\usepackage{ctable}

\newcommand{\stepone}{\scalebox{1.3}{\twemoji{keycap: 1}}}
\newcommand{\steptwo}{\scalebox{1.3}{\twemoji{keycap: 2}}}
\newcommand{\audio}{\scalebox{1.3}{\twemoji{speaker high volume}}}
\newcommand{\texthist}{\scalebox{1.3}{\twemoji{scroll}}}

\definecolor{darkyellow}{RGB}{251,188,4}
\definecolor{darkgreen}{RGB}{52,168,83}
\definecolor{lightblue}{RGB}{66,133,244}
\definecolor{acqua}{RGB}{70,189,198}

\newcommand{\mg}[1]{{\textcolor{black}{#1}}}
\newcommand{\sara}[1]{{\textcolor{black}{#1}}}
\newcommand{\mn}[1]{{\textcolor{black}{#1}}}
\newcommand{\lb}[1]{{\textcolor{black}{#1}}}

\title{\texttt{StreamAtt}: Direct Streaming Speech-to-Text Translation\\with Attention-based Audio History Selection}

\author{Sara Papi \and Marco Gaido  \and  Matteo Negri  \and  Luisa Bentivogli \\
  Fondazione Bruno Kessler, Trento, Italy \\
  \texttt{\{spapi,mgaido,negri,bentivo\}@fbk.eu} \\}

\begin{document}
\maketitle
\begin{abstract}
Streaming speech-to-text translation (StreamST) is the task of automatically translating speech while incrementally receiving an audio stream. Unlike simultaneous ST (SimulST), which deals with pre-segmented speech, StreamST faces the challenges of handling continuous and unbounded audio streams. This requires additional decisions about what to retain of the previous history, which is impractical to keep entirely due to latency and computational constraints. Despite the real-world demand for real-time ST, research on streaming translation remains limited, with existing works solely focusing on SimulST. To fill this gap, we introduce StreamAtt, the first StreamST policy, and propose StreamLAAL, the first StreamST latency metric designed to be comparable with existing metrics for SimulST. Extensive experiments across all 8 languages of MuST-C v1.0 show the effectiveness of StreamAtt compared to a naive streaming baseline and the related state-of-the-art SimulST policy, providing a first step in StreamST research.
\end{abstract}

\section{Introduction}
\label{sec:intro}
Streaming speech-to-text translation (StreamST) is the task of automatically translating spoken content from the source language into the target language 
\sara{in real-time, while continuously receiving an input audio stream. 
By processing longer, unsegmented audio, StreamST} 
\lb{adds another layer of complexity to the difficulties of simultaneous ST (SimulST)
which, instead, operates on -- often manually -- pre-segmented speech segments \citep[among others]{ren-etal-2020-simulspeech,ma-etal-2020-simulmt,liu-etal-2021-cross,weller-etal-2021-streaming,indurthi-etal-2022-language,tang-etal-2023-hybrid}.}

\sara{In SimulST, the primary objective revolves around finding a balance between producing high-quality translations and minimizing latency. This balance}
is managed by a \textbf{simultaneous policy},
which is the strategy for determining, at each time step, whether to emit a partial translation \sara{hypothesis} or to wait for additional audio input. This hypothesis, together with the processed audio, is temporarily stored in memory to provide context for subsequent \sara{generations} and is automatically removed from memory at the end of each audio segment \citep{ma-etal-2020-simuleval}.
\sara{However, when}
the input is a continuous, 
unbounded stream, 
the memory retained as useful context can indefinitely grow, rendering 
\sara{the 
\mn{direct application of} conventional SimulST approaches to StreamST}
impractical due to latency and computational constraints.\footnote{For example, in the SeamlessM4T model for simultaneous translation \citep{barrault2023seamless}, the whole encoder is updated every time a new speech chunk is received (Section 5.2.2 of the paper), which makes its use impracticable for processing continuous, unsegmented audio streams.} 


Despite representing the real-world scenario for providing real-time ST in many applications, such as interpreting \citep{fantinuoli-prandi-2021-towards} and lectures \citep{Fgen2007SimultaneousTO}, and garnering increasing market interests,\footnote{\enquote{The Real-Time Language Translation Device market is anticipated to rise astronomically each year.} (\url{https://www.marketreportsworld.com/enquiry/request-sample/24823921})} research on streaming translation remains limited, with existing works solely focusing on text-to-text machine translation
\lb{(MT)}
\citep{iranzo-sanchez-etal-2022-simultaneous,iranzo2023segmentation}. 
Moreover, as these works focus on (unbounded) text streams as input, there is currently no metric in the literature suitable 
\lb{to evaluate}
the StreamST task, where the input is an audio stream.

\sara{To fill 
these gaps, in this paper we delve into the unexplored domain of StreamST and its associated challenges. First, we define the concept of \textbf{streaming policy} for ST by dividing the decision-making process into two steps: \stepone{} \textit{hypothesis selection}, 
\mn{to determine}
which part of the translation hypothesis should be emitted (akin to the simultaneous policy), and \steptwo{} \textit{history selection},
\mn{to identify}
which part of past audio and generated partial translations should be retained in memory. Then, motivated}
by the success of direct ST models \citep{berard_2016,weiss2017sequence} in overcoming the high latency of cascade architectures in \sara{the related field of} SimulST \citep{Fgen2007SimultaneousTO,fujita13_interspeech,oda-etal-2014-optimizing,muller-etal-2016-lecture,ren-etal-2020-simulspeech}, 
\sara{we propose StreamAtt\footnote{\label{foot:code}Code available at \url{https://github.com/hlt-mt/FBK-fairseq/} under Apache 2.0 license.} (Section \ref{subsec:streamatt}), the first StreamST policy designed for direct ST systems. To enable the evaluation of 
\mn{our}
StreamST solution, we also introduce StreamLAAL\textsuperscript{\ref{foot:code}} (Section \ref{subsec:streamlaal}), 
\mg{the first latency metric for StreamST. StreamLAAL is designed to
\sara{facilitate}
a direct comparison 
with SimulST solutions,
\lb{which} provide upper-bound results as they operate on pre-segmented audio.}
%
%
Lastly, we demonstrate the effectiveness of StreamAtt through extensive experiments across all 8 languages of MuST-C v1.0. We show that 
\mn{our}
policy significantly outperforms a naive streaming baseline (Section \ref{subsec:stream-res}) that relies on a fixed number of past words and audio frames as memory, and is even competitive with the related state-of-the-art SimulST policy at low latency (Section \ref{subsec:simul-comparison}), providing a first promising step in StreamST research.}


\section{Related Works}
While \sara{the terms \enquote{streaming} and \enquote{simultaneous}} translation have often been used interchangeably in the literature, 
\sara{we adhere to}
the definition of streaming by \citet{iranzo-sanchez-etal-2022-simultaneous,iranzo2023segmentation}, 
\mn{which refers to}
the handling of unbounded and continuous streams that, in our case, are audio streams.
Consequently, all works that assume to 
\mn{operate on pre-segmented audio input,}
and only focus on effective methods to determine when and what to emit \citep{ma-etal-2020-simulmt,weller-etal-2021-streaming,liu-etal-2021-cross,indurthi-etal-2022-language,zhang2022information,tang-etal-2023-hybrid}, are hereinafter categorized as related to SimulST.
\mn{Some of these works}
explore how to automatically detect word boundaries in the audio with dedicated modules integrated into the ST system \citep{dong-etal-2022-learning,zhang-etal-2022-learning,fu-etal-2023-adapting,zhang-feng-2023-end}.
However, they still use this information only to determine when and what to emit,
lacking a mechanism to determine which portion of the memory has to be retained and which can be discarded,
hence relying on pre-segmented audio for the simultaneous inference \citep{9414897}.
It follows that all these works are limited to the 
\mn{\textit{hypothesis selection}}
step (i.e., the simultaneous policy) and, differently from this paper, \mn{they all} ignore the \textit{history selection} step, which is necessary to deal with
continuous audio inputs.

\mg{In the context of text-to-text machine translation, the streaming scenario has instead been addressed in \citep{iranzo-sanchez-etal-2022-simultaneous,iranzo2023segmentation}. These works rely on an MT \mn{system} trained with the wait-k simultaneous policy \citep{ma-etal-2019-stacl}, which 
consists 
\mn{in}
waiting for a predefined number of source tokens before starting the translation. The same policy is applied at inference time as the hypothesis selection strategy, while the history selection step consists of keeping a fixed number of textual segments (given by a segmenter model in \citep{iranzo-sanchez-etal-2022-simultaneous} or a memory mechanism integrated into the MT system in \citep{iranzo2023segmentation}) as a context for the current translation generation.}
%
%
%
\mn{As these works are tailored for textual inputs, the evaluation metrics they rely on do not directly apply to StreamST,
\mn{disregarding the crucial aspect of latency measurement from audio streams, which we instead also address in this work.}}

\section{Streaming Policy}
\label{subsec:streamatt}

\begin{figure*}
    \centering
    \includegraphics[width=0.975\textwidth]{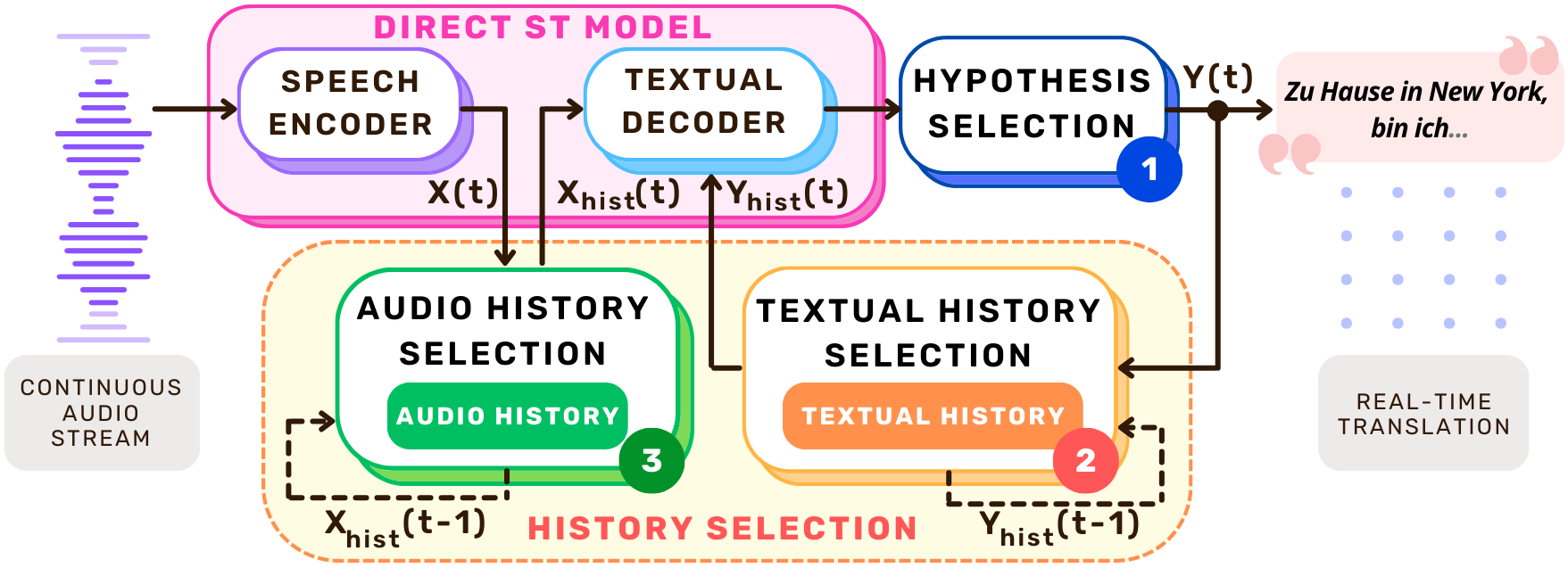}
    \caption{Decision steps of the StreamST policy. The order followed by our StreamAtt policy (step \stepone{}, step \steptwo\texthist, and step \steptwo\audio{}) is indicated from 1 (first) to 3 (last).}
    \label{fig:streamst-policy}
\end{figure*}

The generic decision steps of a StreamST policy (Figure \ref{fig:streamst-policy}) can be schematized as follows:
\begin{enumerate}
    \item[\stepone] \textbf{Hypothesis Selection}: Given both audio and textual history and the newly received chunk of audio, the Hypothesis Selection step 
    determines whether and how many of the newly predicted words to emit.
    This can be easily traced back to the role of a SimulST policy.
    \item[\steptwo] \textbf{History Selection}: Given the audio and textual history retained in the previous step, the newly received speech chunk, and the new partial hypothesis selected in the Hypothesis Selection step, the History Selection step
    decides
    what part of the new history should be retained for processing the next
    audio chunk.
    This process can be further split into:
    \begin{enumerate}
        \item[\texthist] \textit{Textual History Selection}: This sub-step 
        selects the new textual history starting from the textual history retained in the previous 
        iteration
        and the new partial hypothesis obtained by the Hypothesis Selection;
        \item[\audio] \textit{Audio History Selection}: This sub-step 
        selects
        the new audio history starting from the audio history retained in the previous
        iteration
        and the newly received speech chunk.
    \end{enumerate}
\end{enumerate}

Inspired by 
recent findings on the effectiveness of building SimulST systems by 
\mn{directly applying simultaneous policies}
to offline-trained ST models without ad-hoc training/fine-tuning \citep{liu20s_interspeech,Nguyen2021AnES,papi-etal-2022-simultaneous}, 
which 
\mn{led to leadership}
in the IWSLT 2022 Shared Task \citep{anastasopoulos-etal-2022-findings}, 
we 
\mn{develop}
a StreamST policy that exploits
offline-trained ST models.
\mn{In particular, building on recent research proposing cross-attention as a reliable guide for SimulST policies \citep{papi23_interspeech,papi-etal-2023-attention}, we introduce StreamAtt, a StreamST policy that leverages cross-attention scores for both hypothesis and audio history selection (steps \stepone{} and \steptwo\audio), while using a heuristic 
for textual history selection (step \steptwo\texthist).}





\subsection{Hypothesis Selection}
\label{subsec:hypsel}
\mn{For Hypothesis Selection (step \stepone), we exploit \texttt{AlignAtt} \citep{papi23_interspeech},
the state-of-the-art SimulST policy \sara{for offline-trained direct models}.} 
%
%
%
\mn{\texttt{AlignAtt} outperforms all the \sara{alternative} solutions, such as the standard wait-k policy \citep{ma-etal-2019-stacl} adapted for speech either with  \textit{fixed} \citep{ma-etal-2020-simulmt,fukuda-etal-2022-naist,huang-etal-2023-xiaomi} or \textit{adaptive} word boundary detection\footnote{The \textit{fixed} word detection assumes that a word lasts 280ms, while adaptive ones leverage the predictions of a CTC \citep{Graves2006ConnectionistTC} module to determine when a new word starts.} \citep{ren-etal-2020-simulspeech,zeng-etal-2021-realtrans,zeng-etal-2022-end},
and the Local Agreement policy \citep{liu20s_interspeech,polak-etal-2022-cuni}.}
%
%
%
%

\mn{\texttt{AlignAtt}} 
builds upon 
\mn{the observation}
that cross-attention scores can be used 
\mn{to align}
the input and the generated translation \citep{tang-etal-2018-analysis,zenkel2019adding,garg-etal-2019-jointly,chen-etal-2020-accurate},
also with audio as input \citep{papi-etal-2023-attention,alastruey2023speechalign}.
%
Specifically, 
\mg{the alignments between the textual translation $\mathbf{Y}=[y_1,...,y_m]$ and the encoded input audio $\mathbf{X}=[x_1,...,x_n]$ are obtained with the following formula:}
\begin{equation}
\label{eq:align}
    Align(y_i) = \arg \max_{j=1,...,|\mathbf{X}|}  A_{cross}(x_j, y_i) 
\end{equation}
where \mg{$A_{cross}$ 
\mn{stands for}
the cross-attention scores ($A_{cross}$)\footnote{The cross-attention is the dot-product attention \cite{7472621} between the generated tokens $\mathbf{Y}$ and the encoder output $\mathbf{X}$.} computed in the Transformer decoder layers \citep{transformer}.}
$Align(y_i)$ is\mg{, therefore,}
the index of the frame aligned with the predicted token $y_i$\mg{, exploited by \texttt{AlignAtt}}
to decide which tokens of the partial hypothesis have to be emitted. To this aim, it iterates over the predicted tokens and emits them until the following stopping condition is \sara{verified}:
\begin{equation*}
    Align(y_i)>|\mathbf{X}|-f
\end{equation*}
where $f$ is a hyper-parameter that directly controls the latency of the model.
The underlying assumption is that 
if a token is aligned with the most recently received $f$ audio frames, the information provided by these frames can be unstable or insufficiently informative to generate that token (i.e., the system has to wait for additional audio input before generating it).
%
%
%
%
Therefore, smaller $f$ values represent fewer frames 
\mn{that may potentially block the generation}
if attended and, consequently, a lower chance that the stopping condition is verified, resulting in lower latency.


\subsection{History Selection}
%

We design the History Selection of StreamAtt based on two assumptions: \textit{i)} the audio and textual 
history
should be aligned to provide the model with coherent inputs, and \textit{ii)} 
cross-attention
scores provide a 
reliable alignment between the generated text and the input audio, as seen in the previous section. Building on 
these assumptions, we first perform the Textual History Selection 
\steptwo\texthist{} (Section \ref{subsubsec:textsel})
to determine which part of the generated text has to be retained. Then, we forward the resulting textual history to the Audio History Selection 
\steptwo\audio{} (Section \ref{subsubsec:audiosel}), which discards the audio frames that do not align with the provided text. In the following, we 
describe both selection steps.

\subsubsection{Textual History Selection}
\label{subsubsec:textsel}
For Textual History Selection (step \steptwo\texthist), we analyze 
the two different heuristics described below.
\paragraph{Fixed Number of Words (FW).} 
This heuristic retains a fixed number of words ($n_{words}$) in the textual history, inspired by the approach of \citet{iranzo-sanchez-etal-2022-simultaneous} in streaming MT, where a fixed number of segments is retained.
In practice, the textual history $Y_{hist}$ at time $t$ is computed as: 
\begin{equation*}
    Y_{hist}(t) = [Y_{hist}(t-1), Y(t)][:-n_{words}]
\end{equation*}
where $Y(t)$ is the new hypothesis to be emitted at time $t$ and that was selected during step \stepone. Specifically, first, the textual history retained from the previous iteration and the new partial hypothesis are concatenated. Then, only the last $n_{words}$ are preserved as textual history for the next decoding phase (i.e., the next step \stepone).
Since $n_{words}$ is a hyper-parameter, we empirically determined the best value on the validation set, which resulted to be $20$. Detailed results are reported in Appendix \ref{app:fixed-words}.

\paragraph{Punctuation (P).} 
This heuristic simulates what happens in SimulST, where the history is reset at the end of each sentence. 
\mn{As sentence boundaries are not available in StreamST, it considers medium-strong punctuation marks (\enquote{.}, \enquote{!}, \enquote{?}, \enquote{;}, 
\enquote{:}) as sentence boundary proxies. In practice, 
it retains all the words after the last-predicted medium-strong punctuation mark as the textual history $Y_{hist}$ at time $t$. As such, $Y_{hist}$ is computed as:}
\begin{equation*}
    Y_{hist}(t) = [Y_{hist}(t-1), Y(t)][p+1:]
\end{equation*}
\begin{equation*}
     p = \text{last\_index}(\{., !, ?, ;, :\}, [Y_{hist}(t-1), Y(t)])
\end{equation*}
where $Y_{hist}(t-1)$ is the textual history of the previous iteration, $Y(t)$ is the new hypothesis selected in step \stepone, and the function last\_index returns the last occurrence of any of the punctuation marks.
Specifically, only the words after the last punctuation mark are preserved as the textual history for the next step, approximating the reset of the textual history at the end of each sentence as done by SimulST systems.


\subsubsection{Audio History Selection}
\label{subsubsec:audiosel}
For the Audio History Selection (step \steptwo\audio), we exploit the cross-attention scores computed by the model and already used for step \stepone. However, differently from step \stepone{}, these scores are 
used to decide 
which of the currently retained audio frames should be discarded from the audio history.

\mg{First of all, we obtain the current audio $\mathbf{X}$ by concatenating the audio history retained from the previous iteration $X_{hist}(t-1)$ and the new audio input $X(t)$. Then, recalling that we compute the alignment between a textual token and its corresponding audio frame with Eq. \ref{eq:align}, we select the audio history $X_{hist}$ for the iteration $t$ as follows:}
%
\begin{equation*}
    X_{hist}(t) = \mathbf{X}[\min_{h_k\in Y_{hist}(t)}Align(h_k):]
\end{equation*}
Here, $\min_{h_k\in Y_{hist}(t)}Align(h_k)$ represents the index of the first frame of the audio sequence $\mathbf{X}$ that is attended by
at least
one of the tokens $h_k$ in the textual history $Y_{hist}(t)$ determined in the previous step. By doing so, we discard the audio frames that are no longer attended by
the current textual history.
Therefore, the textual history $Y_{hist}(t)$ and audio history $X_{hist}(t)$ preserved at this step, together with the new audio input received in the next step $X(t+1)$, constitute the input of the model for the next iteration.

\section{Streaming Latency Metric}
\label{subsec:streamlaal}

For evaluating the performance of StreamST, the standard metrics adopted in SimulST cannot be applied 
\mn{\textit{as is}.}
In fact,
they are not designed to evaluate outputs obtained from entire audio streams but, instead, refer to (manually) segmented audio and their corresponding translations for the computation. 
\mg{We hence adapt 
\mn{them}
to the streaming scenario to define our StreamST latency metric.}

Specifically, we opt to use the family of latency metrics based on Average Lagging \citep{ma-etal-2019-stacl} for speech \citep{ma-etal-2020-simulmt}, 
given their widespread adoption in SimulST \citep{anastasopoulos-etal-2022-findings,agrawal-etal-2023-findings}. Among these metrics, we select the Length-Adaptive Average Lagging or LAAL \citep{papi-etal-2022-generation,polak-etal-2022-cuni}, which
corrects
the standard AL formulation to avoid the underestimation of the latency when
predictions are longer than the reference translation.
Given $\mathbf{X} = [x_{1}, ..., x_{|\mathbf{X}|}]$ as the speech segment, where each element $x_{j}$ has duration $T_{j}$, $\mathbf{Y^{*}} = [y^{*}_{1}, ..., y^{*}_{|\mathbf{Y^{*}}|}]$ as the reference words,
and $\mathbf{Y} = [y_{1}, ..., y_{|\mathbf{Y}|}]$ as the 
hypothesis
words, LAAL for SimulST is formulated as:
\begin{gather*} 
\label{equation:LAAL}
LAAL = \frac{1}{\tau'(|\mathbf{X}|)}\sum_{i=1}^{\tau'(|\mathbf{X}|)}d_i - d^*_i \\
\label{equation:di}
d^*_i = (i-1) \cdot \frac{\sum_{j=1}^{|\mathbf{X}|} T_j}{\max\{|\mathbf{Y}|,|\mathbf{Y^{*}|\}}}
\end{gather*}
where $d_i$ is the delay of the predicted words, $\tau'(|\mathbf{X}|) =\min\{i|d_i = \sum_{j=1}^{|\mathbf{X}|} T_j\}$ is the index of the target token when the end of the source sentence is reached, and $d^*_i$ represents the delay of an oracle policy that starts to emit words as soon as the speech starts and is perfectly in sync with the speaker.
We
adapt LAAL by considering the entire (unsegmented) stream of audio $\mathbf{S}=[\mathbf{X}_1,...,\mathbf{X}_{|\mathbf{S}|}]$ instead of the single speech segment $\mathbf{X}$, for which we have a continuous stream of predicted translation words $\mathbf{Y_S}$. Since we have reference translations $\mathbf{Y^*_{X_{1}}},...,\mathbf{Y^*_{X_{|\mathbf{S}|}}}$ only for the segmented audio $\mathbf{X_1},...,\mathbf{X}_{|\mathbf{S}|}$, we first obtain the segmented prediction $\mathbf{Y_S}=[\mathbf{Y_{X_1}}, ..., \mathbf{Y_{X_{|\mathbf{S}|}}}]$ with their corresponding delays by applying the mWERSegmenter tool \citep{matusov-etal-2005-evaluating} between each reference sentence $\mathbf{Y^*_i}$ and the entire stream of predicted translation $\mathbf{Y_S}$, in a similar fashion to what has been done for streaming MT \citep{iranzo-sanchez-etal-2021-stream-level}. 
Consequently, we obtain the LAAL for the entire audio stream (StreamLAAL), by computing:
\begin{gather*} 
\label{equation:StreamLAAL}
{Stream \atop LAAL}=\frac{1}{|\mathbf{S}|} \sum_{\mathbf{X_1},...,\mathbf{X_{|\mathbf{S}|}}} \frac{1}{\tau'(|\mathbf{X_i}|)}\sum_{i=1}^{\tau'(|\mathbf{X_i}|)}d_i - d^*_i \\
\label{equation:di}
d^*_i = (i-1) \cdot \frac{\sum_{j=1}^{|\mathbf{X_i}|} T_j}{\max\{|\mathbf{Y_{X_i}}|,|\mathbf{Y^*_{X_i}|\}}}
\end{gather*}
In practice, the LAAL metric is calculated for every speech segment $\mathbf{X_i}$ of the stream $\mathbf{S}$ and its corresponding reference $\mathbf{Y^*_{\mathbf{X_i}}}$ with the automatically aligned prediction $\mathbf{Y_{\mathbf{X_i}}}$ and then averaged over all the speech segments of the stream $\mathbf{X_1},...,\mathbf{X_{|\mathbf{S}|}}$ to obtain StreamLAAL.
\mn{As this formulation builds upon 
the original LAAL metric, it enables direct comparisons between the results obtained in StreamST and those reported in related works on SimulST. In this way, we can 
measure
the gap between StreamST systems and their SimulST 
counterparts, which provide upper-bound results as they operate on pre-segmented audio.}


\section{Experimental Settings}

\subsection{Data}
\label{subsec:data}

To be comparable with previous works \citep{ren-etal-2020-simulspeech,ma-etal-2020-simulmt,zeng-etal-2021-realtrans,chen-etal-2021-direct,liu-etal-2021-cross,zhang2022information,indurthi-etal-2022-language,papi-etal-2022-simultaneous,tang-etal-2023-hybrid}, we train our models on all languages of MuST-C v1.0 \citep{CATTONI2021101155}, namely English (en) to Dutch (nl), French (fr), German (de), Italian (it), Portuguese (pt), Romanian (ro), Russian (ru), and Spanish (es). 

To optimize GPU RAM consumption and speed up training, we filter out segments longer than 30$s$ from the training set.
The resulting data statistics are presented in Table \ref{tab:data}. 

\begin{table}[!htb]
   \centering
   \small
   \setlength{\tabcolsep}{3.5pt}
   \begin{tabular}{cccccccc}
   \specialrule{.1em}{.05em}{.05em} 
       de & es & fr & it & nl & pt & ro & ru \\
       \hline
       225K & 260K & 269K & 248K & 244K & 201K & 231K & 260K \\
   \specialrule{.1em}{.05em}{.05em} 
   \end{tabular}
   \caption{Number of sentences of the training set for each language of MuST-C v1.0.}
   \label{tab:data}
\end{table}

We also perform data augmentation by applying sequence-level knowledge distillation \citep{kim2016sequencelevel,gaido-2020-on-knowledge} as in previous work on SimulST \citep{liu-etal-2021-cross,tang-etal-2023-hybrid}, which consists of translating the transcripts of the training set (MuST-C) with an MT model and using them together with the gold reference during training. 
As a result, the final number of target sentences used during training is twice the original one, while the speech input remains unaltered. 
We use NLLB 3.3B \cite{costa2022no} as the MT model, whose performance on the MuST-C dataset is presented in Appendix \ref{app:nllb}.

\subsection{Architecture and Training Setup}
\label{sec:architecture}

The offline model is a Conformer-based \citep{gulati20_interspeech} encoder-decoder, which is the state-of-the-art architecture in ST \citep{9414858}. All the model details are provided in Appendix \ref{sec:train_setup}.

The input is represented by 80 audio features extracted every 10$ms$ with a sample window of 25 and processed by two 1D Convolutional layers with stride 2 to reduce its length by a factor of 4  \citep{wang2020fairseqs2t}. 
All our models are implemented
in fairseq-s2t \citep{wang2020fairseqs2t}.
Detailed training settings are described in Appendix \ref{sec:train_setup}.

\subsection{Inference, Evaluation, and Comparisons}
\label{subsec:inf-eval-comp}
As StreamAtt is the first StreamST solution, we compare it with a \sara{naive} baseline that \sara{retains} a fixed history both in terms of text and audio. This baseline assumes that each word has a duration of 280$ms$ -- following \citep{ma-etal-2020-simulmt} -- and keeps the same (fixed) number of words in both the audio history and textual history. For the sake of a fair comparison, we set this number of words to $20$, as for StreamAtt (see Section \ref{subsubsec:textsel}).
We also compare StreamAtt with the corresponding state-of-the-art SimulST policy for offline-trained systems \texttt{AlignAtt}.
For both \texttt{AlignAtt} and StreamAtt, we vary the 
hyperparameter $f$ in the range $[2,4,6,8]$ to obtain results for different latency regimes, while we set the size of the speech segment to 1$s$ (the dimension of the incremental speech chunk) and extract the cross-attention scores from the 4\textsuperscript{th} decoder layer, as per \citep{papi-etal-2023-attention}.

We use
our extension of the
SimulEval tool \citep{ma-etal-2020-simuleval} for both SimulST and StreamST evaluation. 
For the streaming approaches (StreamAtt and Baseline), we simulate streaming conditions by providing as input the entire
TED talks of the MuST-C tst-COMMON set.
Instead, for the SimulST \texttt{AlignAtt} policy, we provide the manually segmented audio provided for the same test set, following the standard SimulST evaluation settings. 
We use sacreBLEU \citep{post-2018-call}\footnote{BLEU+case.mixed+smooth.exp+tok.13a+version.2.3.1} for translation quality, and LAAL
-- for \texttt{AlignAtt} -- and StreamLAAL (Section \ref{subsec:streamlaal}) for latency. 
Moreover, as recommended by \citet{ma-etal-2020-simulmt}, we report computationally-aware (CA) StreamLAAL for our streaming comparison, which measures the real elapsed time instead of the ideal latency, as it also accounts for the time required for the model and policy computation.
During inference, the features are computed on the fly and CMVN normalization is based on the global mean and variance estimated on the MuST-C training set. 
Inferences are executed on a NVIDIA K80 GPU with 12GB VRAM.

\begin{table*}[!t]
\setlength{\tabcolsep}{1.5pt}
\small
    \centering
    \begin{tabular}{l||ccc|ccc|ccc|ccc||ccc}
    \specialrule{.1em}{.05em}{.05em}
        \multirow{3}{*}{Strategy} & \multicolumn{3}{c|}{$f=2$} & \multicolumn{3}{c|}{$f=4$} & \multicolumn{3}{c|}{$f=6$} & \multicolumn{3}{c||}{$f=8$} & \multicolumn{3}{c}{\textbf{AVG}} \\
         \cline{2-16}
         & \multirow{2}{*}{BLEU} & \multicolumn{2}{c|}{StreamLAAL} &
         \multirow{2}{*}{BLEU} & \multicolumn{2}{c|}{StreamLAAL} &
         \multirow{2}{*}{BLEU} & \multicolumn{2}{c|}{StreamLAAL} & \multirow{2}{*}{BLEU} & \multicolumn{2}{c||}{StreamLAAL} & \multirow{2}{*}{BLEU} & \multicolumn{2}{c}{StreamLAAL}\\
         \cline{3-4} \cline{6-7} \cline{9-10} \cline{12-13} \cline{15-16}
         & & NCA & CA &  & NCA & CA &  & NCA & CA &  & NCA & CA &   & NCA & CA \\
         \hline
        Baseline & 18.7 & 2.65 & 4.41 & 19.3 & 2.92 & 4.76 & 19.8 & 3.07 & 4.92 & 19.9 & 3.59 & 5.51 & 19.4 & 3.06 & 4.90 \\
        StreamAttFW & 22.3 & \textbf{1.42} & \textbf{2.84} & \textbf{24.3} & \textbf{1.71} & \textbf{3.04} & \textbf{25.1} & \textbf{2.00} & 3.34 & \textbf{25.6} & \textbf{2.30} & \textbf{3.62} & 24.3 & \textbf{1.86} & \textbf{3.21} \\
        StreamAttP & \textbf{22.7} & 1.66 & 3.54 & \textbf{24.3} & 1.84 & 3.81 & 25.0 & 2.15 & \textbf{3.32} & 25.4 & 2.47 & 4.40 & \textbf{24.4} & 2.03 & 3.96 \\
        \specialrule{.1em}{.05em}{.05em}
    \end{tabular}
    \caption{Quality (BLEU$\uparrow$), non-computational and computational aware (NCA/CA) latency (StreamLAAL$\downarrow$) results on MuST-C tst-COMMON averaged over all the 8 languages. Results for each language are shown in Appendix \ref{app:steam-per-lang}.}
    \label{tab:stream-res}
\end{table*}

\section{Results}

In this section, we first compare our proposed StreamAtt policy for StreamST with the streaming baseline (Section \ref{subsec:stream-res}) and then with the state-of-the-art SimulST policy (Section \ref{subsec:simul-comparison}). This 
is followed by an analysis of our approach (Section \ref{subsubsec:punct}).

\subsection{Streaming Results}
\label{subsec:stream-res}

To inspect the streaming ability of the StreamAtt policy equipped either with Fixed Words (FW) or Punctuation (P) Textual History Selection methods (Section \ref{subsubsec:textsel}), we compare its quality-latency performance with a streaming baseline (Section \ref{subsec:inf-eval-comp}). 

The 
translation quality and latency scores,
averaged over the 8 languages of MuST-C v1.0,
are
reported in Table \ref{tab:stream-res}.
Detailed results
for each language pair
can be found in Appendix \ref{app:steam-per-lang}. As can be observed, both StreamAttFW and StreamAttP outperform the baseline by a large margin, with an increase of 5 BLEU points in quality and a reduction of more than 1$s$ in latency, at
every latency regime.
The latency gap further increases when considering the computationally aware latency, with improvements of up to 1.7$s$.
This means that the Audio History Selection strategy of StreamAtt based on cross-attention (Section \ref{subsubsec:audiosel}) is crucial not only for obtaining high-quality translations but also for reducing latency, as discarding audio based solely on fixed duration substantially impacts performance and uselessly increases computational costs.
Furthermore, 
the
significant translation quality drop observed in the baseline
underscores the importance of enforcing alignment between audio and textual history, as StreamAtt does,
and the inadequacy of naive heuristics in \sara{maintaining} this alignment.

Moving to the comparison between StreamAttFW and StreamAttP, the two Textual History Selection methods yield
similar BLEU scores, indicating a similar translation quality. However,
StreamAttFW consistently achieves lower latency both considering computationally unaware 
and aware
latency measures, with 
an average reduction of 170$ms$ in NCA-StreamLAAL and 750$ms$ in CA-StreamLAAL. This result may be surprising, as StreamAttP is designed to mimic the behavior of SimulST systems, but we explain it
in Section \ref{subsubsec:punct}.

\subsection{Comparison with SimulST}
\label{subsec:simul-comparison}
\pgfplotstableread[row sep=\\]{
BLEU	AL \\
19.348  1.255524 \\
22.67  1.432084 \\
24.879  1.667983 \\
26.057  1.909342 \\
}\DEalignatt

\pgfplotstableread[row sep=\\]{
BLEU	AL \\
21.050126506557238       1.497 \\
23.0091355660262        1.806 \\
23.809155301652247      2.061 \\
24.594952864787686      2.361 \\
}\DEstreamatt

\pgfplotstableread[row sep=\\]{
BLEU	AL \\
20.68842587940451      1.525 \\
22.54605733738233       1.861 \\
23.529164185729652      2.126 \\
23.885162192862467       2.433 \\
}\DEstreamattp

\pgfplotstableread[row sep=\\]{
BLEU	AL \\
21.756  1.187064 \\
25.017  1.391932 \\
27.558  1.627583 \\
28.773  1.876933 \\
}\ESalignatt

\pgfplotstableread[row sep=\\]{
BLEU	AL \\
24.437503973344494      1.268 \\
26.22167230405989    1.567 \\
27.281568335747675      1.851 \\
27.812588066816748  2.152 \\
}\ESstreamatt

\pgfplotstableread[row sep=\\]{
BLEU	AL \\
25.179493674874074      1.4887721771799522 \\
26.70487610962769       1.8002123417184239 \\
27.02338687672657       2.0086852920427973 \\
27.56811529559677       2.3917061305778143 \\
}\ESstreamattp

\pgfplotstableread[row sep=\\]{
BLEU	AL \\
27.962  1.205259 \\
32.076  1.377535 \\
35.244  1.612942 \\
36.608  1.861899 \\
}\FRalignatt

\pgfplotstableread[row sep=\\]{
BLEU	AL \\
30.137723185415936      1.456063759393263 \\
32.979423990101836      1.6836522715843516 \\
34.27413842453947       2.0220873977009937 \\
35.040471161681396      2.2629067150013047 \\
}\FRstreamatt

\pgfplotstableread[row sep=\\]{
BLEU	AL \\
31.594967429170577      1.474145043224603 \\
33.589140660916385      1.7447311600003504 \\
34.53719037241545       2.111817333978894 \\
35.14545385865154       2.4131269865331724 \\
}\FRstreamattp

\pgfplotstableread[row sep=\\]{
BLEU	AL \\
19.271  1.268892 \\
22.065  1.450505 \\
24.073  1.689668 \\
25.239  1.932696 \\
}\ITalignatt

\pgfplotstableread[row sep=\\]{
BLEU	AL \\
20.72630547362921       1.3042951894257824 \\
22.60222681944203       1.5763058216113075 \\
23.364289771153388      1.8955214059838147 \\
23.78206914018726       2.215508595492702  \\ 
}\ITstreamatt

\pgfplotstableread[row sep=\\]{
BLEU	AL \\
20.995147614972087      1.5102047733016138 \\
21.725615704437384      1.7144184894265047 \\
23.02011186335794       2.091309829343989 \\
23.17509928569845       2.3888559914676853 \\ 
}\ITstreamattp

\pgfplotstableread[row sep=\\]{
BLEU	AL \\
22.557  1.245954 \\
26.288  1.429317 \\
28.137  1.659568 \\
29.243  1.902384 \\
}\NLalignatt

\pgfplotstableread[row sep=\\]{
BLEU	AL \\
24.01514112543225       1.4683880689072698 \\
26.141307038249682      1.737150312487317 \\
26.879394205866767      2.0024822922220728 \\
27.374099904934933      2.275625765318772 \\
}\NLstreamatt

\pgfplotstableread[row sep=\\]{
BLEU	AL \\
24.111946752293125      1.5879463988049301 \\
25.708771959703615      1.9498311195597964 \\
26.51061952348212       2.1436560819336264 \\
27.074259753307505      2.501414859084182 \\
}\NLstreamattp

\pgfplotstableread[row sep=\\]{
BLEU	AL \\
23.436  1.263355 \\
26.78  1.458153 \\
29.238  1.704334 \\
30.649  1.962534 \\
}\PTalignatt

\pgfplotstableread[row sep=\\]{
BLEU	AL \\
25.54278088983794       1.4322935356637026 \\
27.57718941149826   1.738788287165905 \\
28.276358382590658      2.0087221308847472 \\
28.86065603107929       2.260252189412903 \\
}\PTstreamatt

\pgfplotstableread[row sep=\\]{
BLEU	AL \\
26.36704290445805       1.7232314806596996 \\
27.759248723449534      1.9193099997695554 \\
28.21615281312606       2.200106541246039 \\
28.688774750803724      2.5576844567104736 \\
}\PTstreamattp

\pgfplotstableread[row sep=\\]{
BLEU	AL \\
20.334  1.325662 \\
22.837  1.475423 \\
24.479  1.710009 \\
25.283  1.962288 \\
}\ROalignatt

\pgfplotstableread[row sep=\\]{
BLEU	AL \\
19.62384933140517       1.2888967671924495 \\
21.04270358217957       1.656323936231265 \\
21.667288489371863      1.9388599113674352 \\
22.223005341717066      2.2510747347310007 \\
}\ROstreamatt

\pgfplotstableread[row sep=\\]{
BLEU	AL \\
20.144827492008556      1.5727705632952304 \\
21.61703475491857       1.8810842751541936 \\
22.073184786252153      2.3010694842094417 \\
22.47868767926814       2.515539663801246 \\
}\ROstreamattp

\pgfplotstableread[row sep=\\]{
BLEU	AL \\
12.788  1.349711 \\
14.237  1.450324 \\
15.422  1.655632 \\
16.263  1.901938 \\
}\RUalignatt

\pgfplotstableread[row sep=\\]{
BLEU	AL \\
13.164482700058576      1.650973801442486 \\
14.662378629198768      1.8831747389738869 \\
15.307278564573556      2.2511015310086236 \\
15.406726148775633      2.590761226435057 \\
}\RUstreamatt

\pgfplotstableread[row sep=\\]{
BLEU	AL \\
13.597062990790894      1.558621504529001 \\
14.82528464127627       1.8122513119419295 \\
15.075443014785565      2.22287909967828 \\
15.318269512546358      2.5521479662881584 \\
}\RUstreamattp

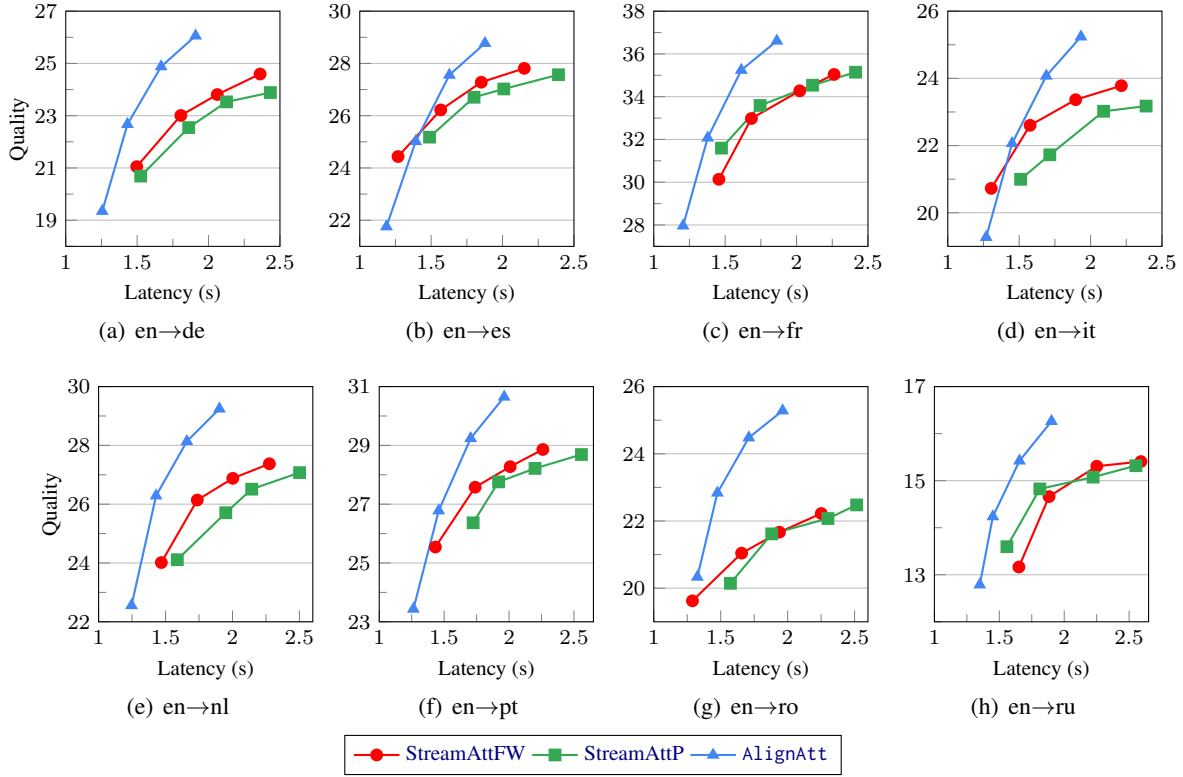
\begin{figure*}[!ht]
\centering
\small
\subfigure[en$\rightarrow$de]{
\begin{tikzpicture}
    \begin{axis}[
            ymajorgrids=true,
            xtick pos=left,
            ytick pos=left,
            minor y tick num=1,
            minor x tick num=1,
            ytick={17,19,21,23,25,27},
            ymin=18,
            ymax=27,
            xmin=1,
            xmax=2.5,
            ylabel=Quality, xlabel=Latency (s),
            ylabel shift={-4pt},
            width=4.4cm,
            height=4.7cm,
            xtick=data,
            compat=newest,
            xtick={1,1.5,2,2.5,3,3.5,4,4.5},
            every axis plot/.append style={thick}
        ]
        \addplot[color=red, mark=*] table[x=AL,y=BLEU]{\DEstreamatt};
        \addplot[color=darkgreen, mark=square*] table[x=AL,y=BLEU]{\DEstreamattp};
        \addplot[color=lightblue, mark=triangle*] table[x=AL,y=BLEU]{\DEalignatt};
    \end{axis}
\end{tikzpicture}
}
\subfigure[en$\rightarrow$es]{
\begin{tikzpicture}
    \footnotesize
    \begin{axis}[
            ymajorgrids=true,
            xtick pos=left,
            ytick pos=left,
            minor y tick num=1,
            minor x tick num=1,
            ytick={22,24,26,28,30},
            ymax=30,
            ymin=21,
            xmin=1,
            xmax=2.5,
            ylabel=, xlabel=Latency (s),
            width=4.4cm,
            height=4.7cm,
            compat=newest,
            xtick=data,
            xtick={1,1.5,2,2.5,3,3.5,4,4.5},
            every axis plot/.append style={thick}
        ]
        \addplot[color=red, mark=*] table[x=AL,y=BLEU]{\ESstreamatt};
        \addplot[color=darkgreen, mark=square*] table[x=AL,y=BLEU]{\ESstreamattp};
        \addplot[color=lightblue, mark=triangle*] table[x=AL,y=BLEU]{\ESalignatt};
    \end{axis}
\end{tikzpicture} 
}
\subfigure[en$\rightarrow$fr]{
\begin{tikzpicture}
    \footnotesize
    \begin{axis}[
            ymajorgrids=true,
            xtick pos=left,
            ytick pos=left,
            minor y tick num=1,
            minor x tick num=1,
            ytick={26,28,30,32,34,36,38},
            ymax=38,
            ymin=27,
            xmin=1,
            xmax=2.5,
            ylabel=, xlabel=Latency (s),
            width=4.4cm,
            height=4.7cm,
            compat=newest,
            xtick=data,
            xtick={1,1.5,2,2.5,3,3.5},
            every axis plot/.append style={thick}
        ]
        \addplot[color=red, mark=*] table[x=AL,y=BLEU]{\FRstreamatt};
        \addplot[color=darkgreen, mark=square*] table[x=AL,y=BLEU]{\FRstreamattp};
        \addplot[color=lightblue, mark=triangle*] table[x=AL,y=BLEU]{\FRalignatt};
    \end{axis}
\end{tikzpicture} 
}
\subfigure[en$\rightarrow$it]{
\begin{tikzpicture}
    \footnotesize
    \begin{axis}[
            ymajorgrids=true,
            xtick pos=left,
            ytick pos=left,
            minor y tick num=1,
            minor x tick num=1,
            ytick={20,22,24,26},
            ymax=26,
            ymin=19,
            xmin=1,
            xmax=2.5,
            ylabel=, xlabel=Latency (s),
            width=4.4cm,
            height=4.7cm,
            compat=newest,
            xtick=data,
            xtick={0.5,1,1.5,2,2.5,3,3.5},
            every axis plot/.append style={thick}
        ]
        \addplot[color=red, mark=*] table[x=AL,y=BLEU]{\ITstreamatt};
        \addplot[color=darkgreen, mark=square*] table[x=AL,y=BLEU]{\ITstreamattp};
        \addplot[color=lightblue, mark=triangle*] table[x=AL,y=BLEU]{\ITalignatt};
    \end{axis}
\end{tikzpicture} 
}
\subfigure[en$\rightarrow$nl]{
\begin{tikzpicture}
    \footnotesize
    \begin{axis}[
            ymajorgrids=true,
            xtick pos=left,
            ytick pos=left,
            minor y tick num=1,
            minor x tick num=1,
            ytick={22,24,26,28,30},
            ymax=30,
            ymin=22,
            xmax=2.6,
            xmin=1,
            ylabel=Quality, xlabel=Latency (s),
            ylabel shift={-4pt},
            width=4.4cm,
            height=4.7cm,
            compat=newest,
            xtick=data,
            xtick={1,1.5,2,2.5,3,3.5},
            every axis plot/.append style={thick}
        ]
        \addplot[color=red, mark=*] table[x=AL,y=BLEU]{\NLstreamatt};
        \addplot[color=darkgreen, mark=square*] table[x=AL,y=BLEU]{\NLstreamattp};
        \addplot[color=lightblue, mark=triangle*] table[x=AL,y=BLEU]{\NLalignatt};
    \end{axis}
\end{tikzpicture} 
}
\subfigure[en$\rightarrow$pt]{
\begin{tikzpicture}
    \footnotesize
    \begin{axis}[
            ymajorgrids=true,
            xtick pos=left,
            ytick pos=left,
            minor y tick num=1,
            minor x tick num=1,
            ytick={23,25,27,29,31},
            ymax=31,
            ymin=23,
            xmin=1,
            xmax=2.65,
            ylabel=, xlabel=Latency (s),
            width=4.4cm,
            height=4.7cm,
            compat=newest,
            xtick=data,
            xtick={1,1.5,2,2.5,3,3.5},
            every axis plot/.append style={thick}
        ]
        \addplot[color=red, mark=*] table[x=AL,y=BLEU]{\PTstreamatt};
        \addplot[color=darkgreen, mark=square*] table[x=AL,y=BLEU]{\PTstreamattp};
        \addplot[color=lightblue, mark=triangle*] table[x=AL,y=BLEU]{\PTalignatt};
    \end{axis}
\end{tikzpicture} 
}
\subfigure[en$\rightarrow$ro]{
\begin{tikzpicture}
    \footnotesize
    \begin{axis}[
            ymajorgrids=true,
            xtick pos=left,
            ytick pos=left,
            minor y tick num=1,
            minor x tick num=1,
            ytick={20,22,24,26},
            ymax=26,
            ymin=19,
            xmin=1,
            xmax=2.6,
            ylabel=, xlabel=Latency (s),
            width=4.4cm,
            height=4.7cm,
            compat=newest,
            xtick=data,
            xtick={1,1.5,2,2.5,3,3.5},
            every axis plot/.append style={thick}
        ]
        \addplot[color=red, mark=*] table[x=AL,y=BLEU]{\ROstreamatt};
        \addplot[color=darkgreen, mark=square*] table[x=AL,y=BLEU]{\ROstreamattp};
        \addplot[color=lightblue, mark=triangle*] table[x=AL,y=BLEU]{\ROalignatt};
    \end{axis}
\end{tikzpicture} 
}
\subfigure[en$\rightarrow$ru]{
\begin{tikzpicture}
    \footnotesize
    \begin{axis}[
            ymajorgrids=true,
            xtick pos=left,
            ytick pos=left,
            minor y tick num=1,
            minor x tick num=1,
            ytick={11,13,15,17},
            ymax=17,
            ymin=12,
            xmin=1,
            xmax=2.65,
            ylabel=, xlabel=Latency (s),
            width=4.4cm,
            height=4.7cm,
            compat=newest,
            xtick=data,
            xtick={1,1.5,2,2.5,3,3.5,4,4.5},
            legend style={at={(0.5,-0.2)},    
                    anchor=north,legend columns=4},   
            legend to name={mylegend},
            every axis plot/.append style={thick}
        ]
        \addplot[color=red, mark=*] table[x=AL,y=BLEU]{\RUstreamatt};
        \addplot[color=darkgreen, mark=square*] table[x=AL,y=BLEU]{\RUstreamattp};
        \addplot[color=lightblue, mark=triangle*] table[x=AL,y=BLEU]{\RUalignatt};
        \legend{StreamAttFW, StreamAttP, \texttt{AlignAtt}}
    \end{axis}
\end{tikzpicture} 
}
\ref{mylegend}
\caption{Latency(LAAL/StreamLAAL$\downarrow$)-Quality(BLEU$\uparrow$) curves of \texttt{AlignAtt} and StreamAtt with Fixed Words (FW) and Punctuation (P) Textual History Selection for all the 8 language pairs of MuST-C v1.0 tst-COMMON.}
\label{fig:simul_res}
\end{figure*}

To further investigate the StreamAtt performance, we compare it with the state-of-the-art \texttt{AlignAtt} policy for SimulST. Since the SimulST policy is applied to manually segmented audio segments, we consider it as an upper-bound for StreamAtt that, instead,
faces the more challenging scenario of unsegmented audio streams. 
Notice that both StreamST and SimulST policies use the same underlying models since they are directly applied to the offline-trained ST systems.

Figure \ref{fig:simul_res} shows the quality-latency plots for each one of the 8 individual languages of MuST-C v1.0. First, it can be observed that, with the only exception of en-ru and en-fr, StreamAttFW \sara{achieves a better quality-latency trade-off compared to StreamAttP}
since the curve of the first is shifted towards the left-upper part compared to the curve of the second. 
Second, we notice that at low latency StreamAtt is close to \texttt{AlignAtt} and is even able to outperform it in some language pairs.
In fact, StreamAttFW yields an improvement of more than 2 BLEU points at 1.2$s$ in Spanish, similar to the gain exhibited in Italian which is of about 1.5 BLEU at the same latency of 1.2$s$. 

Overall, despite being applied to unsegmented speech, the StreamAttFW policy achieves competitive performance at low latency, with less than 1 BLEU of degradation on average across languages, compared to its upper-bound \texttt{AlignAtt}.
Instead, both StreamAttFW and StreamAttP performance is not growing as much as that of \texttt{AlignAtt} when the latency increases, exhibiting a drop of about 2 BLEU points on average. We speculate that the root cause of this behavior (comparable or even better performance at low latency but some quality degradation at higher latency) stems from the intrinsic differences between the simultaneous and streaming tasks:
the simultaneous policy benefits from manually segmented audio, while the streaming policy can use both audio and textual history from previous segments, enhancing performance, especially when this context is more useful, as at lower latency. 
However, when the context becomes too broad, as at higher latency, it can be challenging for the model to effectively select relevant information for the current translation, resulting in performance degradation, as also noted by \citet{iranzo-sanchez-etal-2022-simultaneous} in text-to-text streaming MT.
Further analysis of these aspects presents an interesting avenue for future research.

In summary, despite the added complexities of the streaming task, StreamAtt demonstrates competitive low-latency performance compared to its SimulST upper bound, while closing the quality gap at higher latency is an interesting topic for future StreamST research.

\subsection{Why Punctuation-based Textual History Selection is Worse than Fixed Words?}
\label{subsubsec:punct}

The findings in Sections \ref{subsec:stream-res} and \ref{subsec:simul-comparison} revealed that the streaming solution based on punctuation (StreamAttP) not only exhibits a quality gap at higher latency regimes compared to the SimulST approach 
but also with the fixed words solution (StreamAttFW). 
\sara{To understand this behavior, we carried out a manual inspection of outputs, revealing a noticeable trend across all streaming solutions: they all tend to generate fewer strong punctuation marks, often substituting them with commas.}
To corroborate this observation, we 
computed
the average occurrences of punctuation marks in the outputs of both 
SimulST and StreamST
approaches.

\begin{table}[!htb]
   \centering
   \small
   \setlength{\tabcolsep}{2pt}
   \begin{tabular}{cccccccc}
   \specialrule{.1em}{.05em}{.05em} 
 \textbf{Mark} & \textbf{Reference} & \textbf{SimulST} & \textbf{StreamAttFW} & \textbf{StreamAttP} \\
 \hline
. & 2860.62 & 2651.84 & 1414.34 & 1067.0 \\
! & 10.87 & 2.15 & 1.90 & 1.25 \\
? & 238.87 & 235.09 & 192.65 & 176.37 \\
: & 253.25 & 192.40 & 207.68  & 210.68 \\
; & 49.37 & 10.25 & 26.25 & 24.53 \\
, & 2879.37 & 3835.37 & 5293.62 & 5277.56 \\
   \specialrule{.1em}{.05em}{.05em} 
   \end{tabular}
   \caption{Average number of punctuation marks across all languages and latency regimes for simultaneous and streaming with fixed-words and punctuation-based Textual History Selection compared with the references.}
   \label{tab:punctuation}
\end{table}

As shown in Table \ref{tab:punctuation}, the streaming solutions produce approximately half the number of full stops compared to simultaneous systems (and references), while the occurrence of commas is nearly twice as frequent in streaming outputs compared to simultaneous outputs and references. This 
particular behavior not only sheds light on the underperformance of the punctuation-based solution compared to fixed words (attributed to the scarcity of strong punctuation marks) but also on its quality gap compared to the simultaneous solution.

The 
cause of this issue may be attributed to the fact that systems are trained on manually segmented sentences, which typically feature a single full stop at the end. In the streaming setting, such systems face a mismatch during inference that is, instead, absent in the simultaneous approach executed on audio segmented similarly to the training sets. Given this discovery, an interesting future research direction involves experimenting with data augmentation techniques that introduce samples deviating from the conventional single full-stop placement at the end of speech segments in the training data. For instance, exploring the effects of concatenating multiple sentences in training data \citep{tsz_kin_2023_make} or re-segmenting training data into speech segments that do not correspond to sentences \citep{gaido20_interspeech,lam-etal-2022-sample, tsiamas-etal-2023-segaugment} 
represent promising solutions.


\section{Conclusions}
Our work addressed the
underexplored domain of StreamST, 
which tackles the challenge of translating spoken content from the source language to the target language while incrementally receiving an audio input stream.
Unlike SimulST, which deals with pre-segmented speech chunks, StreamST grapples with the inability to retain the entire growing history in memory due to latency and computational constraints.
Despite growing interest in its applications, research on streaming translation remains limited, with existing studies solely focusing on text-to-text translation, leaving the domain of StreamST and its challenges, including the absence of a suitable evaluation metric, still unaddressed.

\sara{To fill these gaps, in this paper, we delved into the domain of StreamST by first defining the concept of streaming policy for ST. Then, building}
on insights from SimulST research
\sara{underscoring}
the efficacy of direct ST systems in overcoming the latency issues of cascade architectures, we 
\sara{proposed}
StreamAtt, the first StreamST policy tailored for direct ST models\sara{. To enable the evaluation of StreamST solutions, we also introduced}
StreamLAAL, the first StreamST latency metric designed to facilitate direct comparisons with SimulST models.
Through empirical evaluation on all 8 languages of MuST-C v1.0, we showed that StreamAtt significantly outperforms a naive streaming baseline, and is competitive with the SimulST state-of-the-art \texttt{AlignAtt} policy at lower latency, providing a first promising step in StreamST research.

\section*{Acknowledgements}
This paper has received funding from the European Union’s Horizon research and innovation programme under grant agreement No 101135798, project Meetween (My Personal AI Mediator for Virtual MEETtings BetWEEN People).
We also acknowledge the support of the PNRR project FAIR -  Future AI Research (PE00000013),  under the NRRP MUR program funded by the NextGenerationEU. 

% ...

\section*{Limitations}
Although applicable to any offline-trained ST models, StreamAtt and its behavior have been analyzed only on one architectural configuration (12 Conformer encoder layers and 6 Transformer decoder layers). As a consequence, some hyper-parameters, such as the number of words to preserve in the Fixed Words Textual History Selection ($n_{words}$), might vary and depend on the specific ST model, thus requiring a dev set on which to search for the best value before directly testing. Moreover, we applied the Hypothesis Selection-related hyper-parameters (e.g., the number of forbidden frames -- $f$, and the decoder layer from which to extract the cross-attention scores) following previous works but we did not validate these choices on our settings nor changed them to be comparable with these works. Concerning the analyzed languages, the StreamAtt policy has been tested and compared with the naive baseline and related SimulST policy on a restricted set of European languages and, even if there is no reason suggesting that cannot be applied to other languages (possibly after a proper hyper-parameter search), its usage on a wider set of target languages and a source language different from English has not been verified in this work and is left for future research.

As already mentioned in Section \ref{subsubsec:punct}, we have noticed a train-test mismatch between the punctuation of the output emitted by our StreamST policy and the SimulST one, despite both being applied to the same underlying ST model. This underscores that some training or fine-tuning techniques can be applied to further improve StreamAtt performance.
However, besides representing an interesting direction for future research, such investigations were beyond the scope of this study, which aimed to move the first step in the exploration of the StreamST domain.

\bibliography{custom}

\begin{thebibliography}{66}
\expandafter\ifx\csname natexlab\endcsname\relax\def\natexlab#1{#1}\fi

\bibitem[{Agarwal et~al.(2023)Agarwal, Agrawal, Anastasopoulos, Bentivogli, Bojar, Borg, Carpuat, Cattoni, Cettolo, Chen, Chen, Choukri, Chronopoulou, Currey, Declerck, Dong, Duh, Est{\`e}ve, Federico, Gahbiche, Haddow, Hsu, Mon~Htut, Inaguma, Javorsk{\'y}, Judge, Kano, Ko, Kumar, Li, Ma, Mathur, Matusov, McNamee, P.~McCrae, Murray, Nadejde, Nakamura, Negri, Nguyen, Niehues, Niu, Kr.~Ojha, E.~Ortega, Pal, Pino, van~der Plas, Pol{\'a}k, Rippeth, Salesky, Shi, Sperber, St{\"u}ker, Sudoh, Tang, Thompson, Tran, Turchi, Waibel, Wang, Watanabe, and Zevallos}]{agrawal-etal-2023-findings}
Milind Agarwal, Sweta Agrawal, Antonios Anastasopoulos, Luisa Bentivogli, Ond{\v{r}}ej Bojar, Claudia Borg, Marine Carpuat, Roldano Cattoni, Mauro Cettolo, Mingda Chen, William Chen, Khalid Choukri, Alexandra Chronopoulou, Anna Currey, Thierry Declerck, Qianqian Dong, Kevin Duh, Yannick Est{\`e}ve, Marcello Federico, Souhir Gahbiche, Barry Haddow, Benjamin Hsu, Phu Mon~Htut, Hirofumi Inaguma, D{\'a}vid Javorsk{\'y}, John Judge, Yasumasa Kano, Tom Ko, Rishu Kumar, Pengwei Li, Xutai Ma, Prashant Mathur, Evgeny Matusov, Paul McNamee, John P.~McCrae, Kenton Murray, Maria Nadejde, Satoshi Nakamura, Matteo Negri, Ha~Nguyen, Jan Niehues, Xing Niu, Atul Kr.~Ojha, John E.~Ortega, Proyag Pal, Juan Pino, Lonneke van~der Plas, Peter Pol{\'a}k, Elijah Rippeth, Elizabeth Salesky, Jiatong Shi, Matthias Sperber, Sebastian St{\"u}ker, Katsuhito Sudoh, Yun Tang, Brian Thompson, Kevin Tran, Marco Turchi, Alex Waibel, Mingxuan Wang, Shinji Watanabe, and Rodolfo Zevallos. 2023.
\newblock \href {https://doi.org/10.18653/v1/2023.iwslt-1.1} {{FINDINGS} {OF} {THE} {IWSLT} 2023 {EVALUATION} {CAMPAIGN}}.
\newblock In \emph{Proceedings of the 20th International Conference on Spoken Language Translation (IWSLT 2023)}, pages 1--61, Toronto, Canada (in-person and online).

\bibitem[{Alastruey et~al.(2023)Alastruey, Sant, G{\'a}llego, Dale, and Costa-juss{\`a}}]{alastruey2023speechalign}
Belen Alastruey, Aleix Sant, Gerard~I G{\'a}llego, David Dale, and Marta~R Costa-juss{\`a}. 2023.
\newblock Speechalign: a framework for speech translation alignment evaluation.
\newblock \emph{arXiv preprint arXiv:2309.11585}.

\bibitem[{Anastasopoulos et~al.(2022)Anastasopoulos, Barrault, Bentivogli, Zanon~Boito, Bojar, Cattoni, Currey, Dinu, Duh, Elbayad, Emmanuel, Est{\`e}ve, Federico, Federmann, Gahbiche, Gong, Grundkiewicz, Haddow, Hsu, Javorsk{\'y}, Kloudov{\'a}, Lakew, Ma, Mathur, McNamee, Murray, N{\v{a}}dejde, Nakamura, Negri, Niehues, Niu, Ortega, Pino, Salesky, Shi, Sperber, St{\"u}ker, Sudoh, Turchi, Virkar, Waibel, Wang, and Watanabe}]{anastasopoulos-etal-2022-findings}
Antonios Anastasopoulos, Lo{\"\i}c Barrault, Luisa Bentivogli, Marcely Zanon~Boito, Ond{\v{r}}ej Bojar, Roldano Cattoni, Anna Currey, Georgiana Dinu, Kevin Duh, Maha Elbayad, Clara Emmanuel, Yannick Est{\`e}ve, Marcello Federico, Christian Federmann, Souhir Gahbiche, Hongyu Gong, Roman Grundkiewicz, Barry Haddow, Benjamin Hsu, D{\'a}vid Javorsk{\'y}, V{\u{e}}ra Kloudov{\'a}, Surafel Lakew, Xutai Ma, Prashant Mathur, Paul McNamee, Kenton Murray, Maria N{\v{a}}dejde, Satoshi Nakamura, Matteo Negri, Jan Niehues, Xing Niu, John Ortega, Juan Pino, Elizabeth Salesky, Jiatong Shi, Matthias Sperber, Sebastian St{\"u}ker, Katsuhito Sudoh, Marco Turchi, Yogesh Virkar, Alexander Waibel, Changhan Wang, and Shinji Watanabe. 2022.
\newblock \href {https://doi.org/10.18653/v1/2022.iwslt-1.10} {Findings of the {IWSLT} 2022 evaluation campaign}.
\newblock In \emph{Proceedings of the 19th International Conference on Spoken Language Translation (IWSLT 2022)}, pages 98--157, Dublin, Ireland (in-person and online).

\bibitem[{Barrault et~al.(2023)Barrault, Chung, Meglioli, Dale, Dong, Duppenthaler, Duquenne, Ellis, Elsahar, Haaheim et~al.}]{barrault2023seamless}
Lo{\"\i}c Barrault, Yu-An Chung, Mariano~Coria Meglioli, David Dale, Ning Dong, Mark Duppenthaler, Paul-Ambroise Duquenne, Brian Ellis, Hady Elsahar, Justin Haaheim, et~al. 2023.
\newblock Seamless: Multilingual expressive and streaming speech translation.
\newblock \emph{arXiv preprint arXiv:2312.05187}.

\bibitem[{Bérard et~al.(2016)Bérard, Pietquin, Servan, and Besacier}]{berard_2016}
Alexandre Bérard, Olivier Pietquin, Christophe Servan, and Laurent Besacier. 2016.
\newblock {Listen and Translate: A Proof of Concept for End-to-End Speech-to-Text Translation}.
\newblock In \emph{NIPS Workshop on end-to-end learning for speech and audio processing}, Barcelona, Spain.

\bibitem[{Cattoni et~al.(2021)Cattoni, {Di Gangi}, Bentivogli, Negri, and Turchi}]{CATTONI2021101155}
Roldano Cattoni, Mattia~Antonino {Di Gangi}, Luisa Bentivogli, Matteo Negri, and Marco Turchi. 2021.
\newblock \href {https://doi.org/https://doi.org/10.1016/j.csl.2020.101155} {Must-c: A multilingual corpus for end-to-end speech translation}.
\newblock \emph{Computer Speech \& Language}, 66:101155.

\bibitem[{Chan et~al.(2016)Chan, Jaitly, Le, and Vinyals}]{7472621}
William Chan, Navdeep Jaitly, Quoc Le, and Oriol Vinyals. 2016.
\newblock \href {https://doi.org/10.1109/ICASSP.2016.7472621} {Listen, attend and spell: A neural network for large vocabulary conversational speech recognition}.
\newblock In \emph{2016 IEEE International Conference on Acoustics, Speech and Signal Processing (ICASSP)}, pages 4960--4964.

\bibitem[{Chen et~al.(2021)Chen, Ma, Zheng, and Huang}]{chen-etal-2021-direct}
Junkun Chen, Mingbo Ma, Renjie Zheng, and Liang Huang. 2021.
\newblock \href {https://doi.org/10.18653/v1/2021.findings-acl.406} {Direct simultaneous speech-to-text translation assisted by synchronized streaming {ASR}}.
\newblock In \emph{Findings of the Association for Computational Linguistics: ACL-IJCNLP 2021}, pages 4618--4624, Online.

\bibitem[{Chen et~al.(2020)Chen, Liu, Chen, Jiang, and Liu}]{chen-etal-2020-accurate}
Yun Chen, Yang Liu, Guanhua Chen, Xin Jiang, and Qun Liu. 2020.
\newblock \href {https://doi.org/10.18653/v1/2020.emnlp-main.42} {Accurate word alignment induction from neural machine translation}.
\newblock In \emph{Proceedings of the 2020 Conference on Empirical Methods in Natural Language Processing (EMNLP)}, pages 566--576, Online.

\bibitem[{Costa-juss{\`a} et~al.(2022)Costa-juss{\`a}, Cross, {\c{C}}elebi, Elbayad, Heafield, Heffernan, Kalbassi, Lam, Licht, Maillard et~al.}]{costa2022no}
Marta~R Costa-juss{\`a}, James Cross, Onur {\c{C}}elebi, Maha Elbayad, Kenneth Heafield, Kevin Heffernan, Elahe Kalbassi, Janice Lam, Daniel Licht, Jean Maillard, et~al. 2022.
\newblock No language left behind: Scaling human-centered machine translation.
\newblock \emph{arXiv preprint arXiv:2207.04672}.

\bibitem[{Dong et~al.(2022)Dong, Zhu, Wang, and Li}]{dong-etal-2022-learning}
Qian Dong, Yaoming Zhu, Mingxuan Wang, and Lei Li. 2022.
\newblock \href {https://doi.org/10.18653/v1/2022.acl-long.50} {Learning when to translate for streaming speech}.
\newblock In \emph{Proceedings of the 60th Annual Meeting of the Association for Computational Linguistics (Volume 1: Long Papers)}, pages 680--694, Dublin, Ireland.

\bibitem[{Fantinuoli and Prandi(2021)}]{fantinuoli-prandi-2021-towards}
Claudio Fantinuoli and Bianca Prandi. 2021.
\newblock \href {https://doi.org/10.18653/v1/2021.iwslt-1.29} {Towards the evaluation of automatic simultaneous speech translation from a communicative perspective}.
\newblock In \emph{Proceedings of the 18th International Conference on Spoken Language Translation (IWSLT 2021)}, pages 245--254, Bangkok, Thailand (online). Association for Computational Linguistics.

\bibitem[{Fu et~al.(2023)Fu, Liao, Fan, Huang, Chen, Chen, and Shi}]{fu-etal-2023-adapting}
Biao Fu, Minpeng Liao, Kai Fan, Zhongqiang Huang, Boxing Chen, Yidong Chen, and Xiaodong Shi. 2023.
\newblock \href {https://doi.org/10.18653/v1/2023.emnlp-main.1033} {Adapting offline speech translation models for streaming with future-aware distillation and inference}.
\newblock In \emph{Proceedings of the 2023 Conference on Empirical Methods in Natural Language Processing}, pages 16600--16619, Singapore.

\bibitem[{F{\"u}gen et~al.(2007)F{\"u}gen, Waibel, and Kolss}]{Fgen2007SimultaneousTO}
Christian F{\"u}gen, Alexander~H. Waibel, and Muntsin Kolss. 2007.
\newblock Simultaneous translation of lectures and speeches.
\newblock \emph{Machine Translation}, 21:209--252.

\bibitem[{Fujita et~al.(2013)Fujita, Neubig, Sakti, Toda, and Nakamura}]{fujita13_interspeech}
Tomoki Fujita, Graham Neubig, Sakriani Sakti, Tomoki Toda, and Satoshi Nakamura. 2013.
\newblock \href {https://doi.org/10.21437/Interspeech.2013-615} {{Simple, lexicalized choice of translation timing for simultaneous speech translation}}.
\newblock In \emph{Proc. Interspeech 2013}, pages 3487--3491.

\bibitem[{Fukuda et~al.(2022)Fukuda, Ko, Kano, Doi, Tokuyama, Sakti, Sudoh, and Nakamura}]{fukuda-etal-2022-naist}
Ryo Fukuda, Yuka Ko, Yasumasa Kano, Kosuke Doi, Hirotaka Tokuyama, Sakriani Sakti, Katsuhito Sudoh, and Satoshi Nakamura. 2022.
\newblock \href {https://doi.org/10.18653/v1/2022.iwslt-1.25} {{NAIST} simultaneous speech-to-text translation system for {IWSLT} 2022}.
\newblock In \emph{Proceedings of the 19th International Conference on Spoken Language Translation (IWSLT 2022)}, pages 286--292, Dublin, Ireland (in-person and online).

\bibitem[{Gaido et~al.(2021{\natexlab{a}})Gaido, Cettolo, Negri, and Turchi}]{gaido-etal-2021-ctc}
Marco Gaido, Mauro Cettolo, Matteo Negri, and Marco Turchi. 2021{\natexlab{a}}.
\newblock \href {https://doi.org/10.18653/v1/2021.eacl-main.57} {{CTC}-based compression for direct speech translation}.
\newblock In \emph{Proceedings of the 16th Conference of the European Chapter of the Association for Computational Linguistics: Main Volume}, pages 690--696, Online.

\bibitem[{Gaido et~al.(2021{\natexlab{b}})Gaido, Di~Gangi, Negri, and Turchi}]{gaido-2020-on-knowledge}
Marco Gaido, Mattia~A. Di~Gangi, Matteo Negri, and Marco Turchi. 2021{\natexlab{b}}.
\newblock \href {http://ceur-ws.org/Vol-2769/paper_28.pdf} {{On Knowledge Distillation for Direct Speech Translation }}.
\newblock In \emph{Proceedings of CLiC-IT 2020}, Online.

\bibitem[{Gaido et~al.(2020)Gaido, Gangi, Negri, Cettolo, and Turchi}]{gaido20_interspeech}
Marco Gaido, Mattia A.~Di Gangi, Matteo Negri, Mauro Cettolo, and Marco Turchi. 2020.
\newblock \href {https://doi.org/10.21437/Interspeech.2020-2860} {{Contextualized Translation of Automatically Segmented Speech}}.
\newblock In \emph{Proc. Interspeech 2020}, pages 1471--1475.

\bibitem[{Gaido et~al.(2022)Gaido, Papi, Fucci, Fiameni, Negri, and Turchi}]{gaido-etal-2022-efficient}
Marco Gaido, Sara Papi, Dennis Fucci, Giuseppe Fiameni, Matteo Negri, and Marco Turchi. 2022.
\newblock \href {https://doi.org/10.18653/v1/2022.iwslt-1.13} {Efficient yet competitive speech translation: {FBK}@{IWSLT}2022}.
\newblock In \emph{Proceedings of the 19th International Conference on Spoken Language Translation (IWSLT 2022)}, pages 177--189, Dublin, Ireland (in-person and online).

\bibitem[{Garg et~al.(2019)Garg, Peitz, Nallasamy, and Paulik}]{garg-etal-2019-jointly}
Sarthak Garg, Stephan Peitz, Udhyakumar Nallasamy, and Matthias Paulik. 2019.
\newblock \href {https://doi.org/10.18653/v1/D19-1453} {Jointly learning to align and translate with transformer models}.
\newblock In \emph{Proceedings of the 2019 Conference on Empirical Methods in Natural Language Processing and the 9th International Joint Conference on Natural Language Processing (EMNLP-IJCNLP)}, pages 4453--4462, Hong Kong, China.

\bibitem[{Graves et~al.(2006)Graves, Fern{\'a}ndez, Gomez, and Schmidhuber}]{Graves2006ConnectionistTC}
Alex Graves, Santiago Fern{\'a}ndez, Faustino~J. Gomez, and J{\"u}rgen Schmidhuber. 2006.
\newblock {Connectionist Temporal Classification: Labelling Unsegmented Sequence Data with Recurrent Neural Networks}.
\newblock In \emph{Proceedings of the 23rd international conference on Machine learning (ICML)}, pages 369--376, Pittsburgh, Pennsylvania.

\bibitem[{Gulati et~al.(2020)Gulati, Qin, Chiu, Parmar, Zhang, Yu, Han, Wang, Zhang, Wu, and Pang}]{gulati20_interspeech}
Anmol Gulati, James Qin, Chung-Cheng Chiu, Niki Parmar, Yu~Zhang, Jiahui Yu, Wei Han, Shibo Wang, Zhengdong Zhang, Yonghui Wu, and Ruoming Pang. 2020.
\newblock \href {https://doi.org/10.21437/Interspeech.2020-3015} {{Conformer: Convolution-augmented Transformer for Speech Recognition}}.
\newblock In \emph{Proc. Interspeech 2020}, pages 5036--5040.

\bibitem[{Guo et~al.(2021)Guo, Boyer, Chang, Hayashi, Higuchi, Inaguma, Kamo, Li, Garcia-Romero, Shi, Shi, Watanabe, Wei, Zhang, and Zhang}]{9414858}
Pengcheng Guo, Florian Boyer, Xuankai Chang, Tomoki Hayashi, Yosuke Higuchi, Hirofumi Inaguma, Naoyuki Kamo, Chenda Li, Daniel Garcia-Romero, Jiatong Shi, Jing Shi, Shinji Watanabe, Kun Wei, Wangyou Zhang, and Yuekai Zhang. 2021.
\newblock \href {https://doi.org/10.1109/ICASSP39728.2021.9414858} {Recent developments on espnet toolkit boosted by conformer}.
\newblock In \emph{ICASSP 2021 - 2021 IEEE International Conference on Acoustics, Speech and Signal Processing (ICASSP)}, pages 5874--5878.

\bibitem[{Huang et~al.(2023)Huang, Liu, Li, Tian, Yang, Zhang, Luan, Wang, Guo, and Su}]{huang-etal-2023-xiaomi}
Wuwei Huang, Mengge Liu, Xiang Li, Yanzhi Tian, Fengyu Yang, Wen Zhang, Jian Luan, Bin Wang, Yuhang Guo, and Jinsong Su. 2023.
\newblock \href {https://doi.org/10.18653/v1/2023.iwslt-1.39} {The xiaomi {AI} lab{'}s speech translation systems for {IWSLT} 2023 offline task, simultaneous task and speech-to-speech task}.
\newblock In \emph{Proceedings of the 20th International Conference on Spoken Language Translation (IWSLT 2023)}, pages 411--419, Toronto, Canada (in-person and online). Association for Computational Linguistics.

\bibitem[{Indurthi et~al.(2022)Indurthi, Zaidi, Lee, Lakumarapu, and Kim}]{indurthi-etal-2022-language}
Sathish~Reddy Indurthi, Mohd~Abbas Zaidi, Beomseok Lee, Nikhil~Kumar Lakumarapu, and Sangha Kim. 2022.
\newblock \href {https://doi.org/10.18653/v1/2022.naacl-main.3} {Language model augmented monotonic attention for simultaneous translation}.
\newblock In \emph{Proceedings of the 2022 Conference of the North American Chapter of the Association for Computational Linguistics: Human Language Technologies}, pages 38--45, Seattle, United States.

\bibitem[{Iranzo-S{\'a}nchez et~al.(2022)Iranzo-S{\'a}nchez, Civera, and Juan}]{iranzo-sanchez-etal-2022-simultaneous}
Javier Iranzo-S{\'a}nchez, Jorge Civera, and Alfons Juan. 2022.
\newblock \href {https://doi.org/10.18653/v1/2022.acl-long.480} {From simultaneous to streaming machine translation by leveraging streaming history}.
\newblock In \emph{Proceedings of the 60th Annual Meeting of the Association for Computational Linguistics (Volume 1: Long Papers)}, pages 6972--6985, Dublin, Ireland. Association for Computational Linguistics.

\bibitem[{Iranzo-S{\'a}nchez et~al.(2021)Iranzo-S{\'a}nchez, Civera~Saiz, and Juan}]{iranzo-sanchez-etal-2021-stream-level}
Javier Iranzo-S{\'a}nchez, Jorge Civera~Saiz, and Alfons Juan. 2021.
\newblock \href {https://doi.org/10.18653/v1/2021.findings-emnlp.58} {Stream-level latency evaluation for simultaneous machine translation}.
\newblock In \emph{Findings of the Association for Computational Linguistics: EMNLP 2021}, pages 664--670, Punta Cana, Dominican Republic. Association for Computational Linguistics.

\bibitem[{Iranzo-S{\'a}nchez et~al.(2023)Iranzo-S{\'a}nchez, Iranzo-S{\'a}nchez, Gim{\'e}nez, Civera, and Juan}]{iranzo2023segmentation}
Javier Iranzo-S{\'a}nchez, Jorge Iranzo-S{\'a}nchez, Adri{\`a} Gim{\'e}nez, Jorge Civera, and Alfons Juan. 2023.
\newblock Segmentation-free streaming machine translation.
\newblock \emph{arXiv preprint arXiv:2309.14823}.

\bibitem[{Kim and Rush(2016)}]{kim2016sequencelevel}
Yoon Kim and Alexander~M. Rush. 2016.
\newblock \href {https://doi.org/10.18653/v1/D16-1139} {{Sequence-Level Knowledge Distillation}}.
\newblock In \emph{Proc. of the 2016 Conference on Empirical Methods in Natural Language Processing}, pages 1317--1327, Austin, Texas.

\bibitem[{Kingma and Ba(2015)}]{DBLP:journals/corr/KingmaB14}
Diederik~P. Kingma and Jimmy Ba. 2015.
\newblock \href {http://arxiv.org/abs/1412.6980} {Adam: {A} method for stochastic optimization}.
\newblock In \emph{3rd International Conference on Learning Representations, {ICLR} 2015, San Diego, CA, USA, May 7-9, 2015, Conference Track Proceedings}.

\bibitem[{Kudo and Richardson(2018)}]{kudo-richardson-2018-sentencepiece}
Taku Kudo and John Richardson. 2018.
\newblock \href {https://doi.org/10.18653/v1/D18-2012} {{S}entence{P}iece: A simple and language independent subword tokenizer and detokenizer for neural text processing}.
\newblock In \emph{Proceedings of the 2018 Conference on Empirical Methods in Natural Language Processing: System Demonstrations}, pages 66--71, Brussels, Belgium. Association for Computational Linguistics.

\bibitem[{Lam et~al.(2022)Lam, Schamoni, and Riezler}]{lam-etal-2022-sample}
Tsz~Kin Lam, Shigehiko Schamoni, and Stefan Riezler. 2022.
\newblock \href {https://doi.org/10.18653/v1/2022.acl-short.27} {Sample, translate, recombine: Leveraging audio alignments for data augmentation in end-to-end speech translation}.
\newblock In \emph{Proceedings of the 60th Annual Meeting of the Association for Computational Linguistics (Volume 2: Short Papers)}, pages 245--254, Dublin, Ireland. Association for Computational Linguistics.

\bibitem[{Lam et~al.(2023)Lam, Schamoni, and Riezler}]{tsz_kin_2023_make}
Tsz~Kin Lam, Shigehiko Schamoni, and Stefan Riezler. 2023.
\newblock \href {https://doi.org/10.1109/ICASSP49357.2023.10094564} {Make more of your data: Minimal effort data augmentation for automatic speech recognition and translation}.
\newblock In \emph{ICASSP 2023 - 2023 IEEE International Conference on Acoustics, Speech and Signal Processing (ICASSP)}, pages 1--5.

\bibitem[{Liu et~al.(2021)Liu, Du, Li, Li, and Chen}]{liu-etal-2021-cross}
Dan Liu, Mengge Du, Xiaoxi Li, Ya~Li, and Enhong Chen. 2021.
\newblock \href {https://aclanthology.org/2021.emnlp-main.4} {Cross attention augmented transducer networks for simultaneous translation}.
\newblock In \emph{Proceedings of the 2021 Conference on Empirical Methods in Natural Language Processing}, pages 39--55, Online and Punta Cana, Dominican Republic.

\bibitem[{Liu et~al.(2020)Liu, Spanakis, and Niehues}]{liu20s_interspeech}
Danni Liu, Gerasimos Spanakis, and Jan Niehues. 2020.
\newblock \href {https://doi.org/10.21437/Interspeech.2020-2897} {{Low-Latency Sequence-to-Sequence Speech Recognition and Translation by Partial Hypothesis Selection}}.
\newblock In \emph{Proc. Interspeech 2020}, pages 3620--3624.

\bibitem[{Ma et~al.(2019)Ma, Huang, Xiong, Zheng, Liu, Zheng, Zhang, He, Liu, Li, Wu, and Wang}]{ma-etal-2019-stacl}
Mingbo Ma, Liang Huang, Hao Xiong, Renjie Zheng, Kaibo Liu, Baigong Zheng, Chuanqiang Zhang, Zhongjun He, Hairong Liu, Xing Li, Hua Wu, and Haifeng Wang. 2019.
\newblock \href {https://doi.org/10.18653/v1/P19-1289} {{STACL}: Simultaneous translation with implicit anticipation and controllable latency using prefix-to-prefix framework}.
\newblock In \emph{Proceedings of the 57th Annual Meeting of the Association for Computational Linguistics}, pages 3025--3036, Florence, Italy.

\bibitem[{Ma et~al.(2020{\natexlab{a}})Ma, Dousti, Wang, Gu, and Pino}]{ma-etal-2020-simuleval}
Xutai Ma, Mohammad~Javad Dousti, Changhan Wang, Jiatao Gu, and Juan Pino. 2020{\natexlab{a}}.
\newblock \href {https://doi.org/10.18653/v1/2020.emnlp-demos.19} {{SIMULEVAL}: An evaluation toolkit for simultaneous translation}.
\newblock In \emph{Proceedings of the 2020 Conference on Empirical Methods in Natural Language Processing: System Demonstrations}, pages 144--150, Online.

\bibitem[{Ma et~al.(2020{\natexlab{b}})Ma, Pino, and Koehn}]{ma-etal-2020-simulmt}
Xutai Ma, Juan Pino, and Philipp Koehn. 2020{\natexlab{b}}.
\newblock \href {https://www.aclweb.org/anthology/2020.aacl-main.58} {{S}imul{MT} to {S}imul{ST}: Adapting simultaneous text translation to end-to-end simultaneous speech translation}.
\newblock In \emph{Proceedings of the 1st Conference of the Asia-Pacific Chapter of the Association for Computational Linguistics and the 10th International Joint Conference on Natural Language Processing}, pages 582--587, Suzhou, China.

\bibitem[{Ma et~al.(2021)Ma, Wang, Dousti, Koehn, and Pino}]{9414897}
Xutai Ma, Yongqiang Wang, Mohammad~Javad Dousti, Philipp Koehn, and Juan Pino. 2021.
\newblock \href {https://doi.org/10.1109/ICASSP39728.2021.9414897} {Streaming simultaneous speech translation with augmented memory transformer}.
\newblock In \emph{ICASSP 2021 - 2021 IEEE International Conference on Acoustics, Speech and Signal Processing (ICASSP)}, pages 7523--7527.

\bibitem[{Matusov et~al.(2005)Matusov, Leusch, Bender, and Ney}]{matusov-etal-2005-evaluating}
Evgeny Matusov, Gregor Leusch, Oliver Bender, and Hermann Ney. 2005.
\newblock \href {https://aclanthology.org/2005.iwslt-1.19} {Evaluating machine translation output with automatic sentence segmentation}.
\newblock In \emph{Proceedings of the Second International Workshop on Spoken Language Translation}, Pittsburgh, Pennsylvania, USA.

\bibitem[{M{\"u}ller et~al.(2016)M{\"u}ller, Nguyen, Niehues, Cho, Kr{\"u}ger, Ha, Kilgour, Sperber, Mediani, St{\"u}ker, and Waibel}]{muller-etal-2016-lecture}
Markus M{\"u}ller, Thai~Son Nguyen, Jan Niehues, Eunah Cho, Bastian Kr{\"u}ger, Thanh-Le Ha, Kevin Kilgour, Matthias Sperber, Mohammed Mediani, Sebastian St{\"u}ker, and Alex Waibel. 2016.
\newblock \href {https://doi.org/10.18653/v1/N16-3017} {Lecture translator - speech translation framework for simultaneous lecture translation}.
\newblock In \emph{Proceedings of the 2016 Conference of the North {A}merican Chapter of the Association for Computational Linguistics: Demonstrations}, pages 82--86, San Diego, California. Association for Computational Linguistics.

\bibitem[{Nguyen et~al.(2021)Nguyen, Est{\`e}ve, and Besacier}]{Nguyen2021AnES}
Ha~Nguyen, Y.~Est{\`e}ve, and Laurent Besacier. 2021.
\newblock An empirical study of end-to-end simultaneous speech translation decoding strategies.
\newblock \emph{ICASSP 2021 - 2021 IEEE International Conference on Acoustics, Speech and Signal Processing (ICASSP)}, pages 7528--7532.

\bibitem[{Oda et~al.(2014)Oda, Neubig, Sakti, Toda, and Nakamura}]{oda-etal-2014-optimizing}
Yusuke Oda, Graham Neubig, Sakriani Sakti, Tomoki Toda, and Satoshi Nakamura. 2014.
\newblock \href {https://doi.org/10.3115/v1/P14-2090} {Optimizing segmentation strategies for simultaneous speech translation}.
\newblock In \emph{Proceedings of the 52nd Annual Meeting of the Association for Computational Linguistics (Volume 2: Short Papers)}, pages 551--556, Baltimore, Maryland. Association for Computational Linguistics.

\bibitem[{Papi et~al.(2022{\natexlab{a}})Papi, Gaido, Negri, and Turchi}]{papi-etal-2022-simultaneous}
Sara Papi, Marco Gaido, Matteo Negri, and Marco Turchi. 2022{\natexlab{a}}.
\newblock \href {https://doi.org/10.18653/v1/2022.findings-emnlp.11} {Does simultaneous speech translation need simultaneous models?}
\newblock In \emph{Findings of the Association for Computational Linguistics: EMNLP 2022}, pages 141--153, Abu Dhabi, United Arab Emirates. Association for Computational Linguistics.

\bibitem[{Papi et~al.(2022{\natexlab{b}})Papi, Gaido, Negri, and Turchi}]{papi-etal-2022-generation}
Sara Papi, Marco Gaido, Matteo Negri, and Marco Turchi. 2022{\natexlab{b}}.
\newblock \href {https://doi.org/10.18653/v1/2022.autosimtrans-1.2} {Over-generation cannot be rewarded: Length-adaptive average lagging for simultaneous speech translation}.
\newblock In \emph{Proceedings of the Third Workshop on Automatic Simultaneous Translation}, pages 12--17, Online.

\bibitem[{Papi et~al.(2023{\natexlab{a}})Papi, Negri, and Turchi}]{papi-etal-2023-attention}
Sara Papi, Matteo Negri, and Marco Turchi. 2023{\natexlab{a}}.
\newblock \href {https://doi.org/10.18653/v1/2023.acl-long.745} {{Attention as a Guide for Simultaneous Speech Translation}}.
\newblock In \emph{Proceedings of the 61st Annual Meeting of the Association for Computational Linguistics (Volume 1: Long Papers)}, pages 13340--13356, Toronto, Canada.

\bibitem[{Papi et~al.(2023{\natexlab{b}})Papi, Turchi, and Negri}]{papi23_interspeech}
Sara Papi, Marco Turchi, and Matteo Negri. 2023{\natexlab{b}}.
\newblock \href {https://doi.org/10.21437/Interspeech.2023-170} {{AlignAtt: Using Attention-based Audio-Translation Alignments as a Guide for Simultaneous Speech Translation}}.
\newblock In \emph{Proc. INTERSPEECH 2023}, pages 3974--3978.

\bibitem[{Park et~al.(2019)Park, Chan, Zhang, Chiu, Zoph, Cubuk, and Le}]{Park2019}
Daniel~S. Park, William Chan, Yu~Zhang, Chung-Cheng Chiu, Barret Zoph, Ekin~D. Cubuk, and Quoc~V. Le. 2019.
\newblock \href {https://doi.org/10.21437/Interspeech.2019-2680} {{SpecAugment: A Simple Data Augmentation Method for Automatic Speech Recognition}}.
\newblock In \emph{Proc. Interspeech 2019}, pages 2613--2617.

\bibitem[{Pol{\'a}k et~al.(2022)Pol{\'a}k, Pham, Nguyen, Liu, Mullov, Niehues, Bojar, and Waibel}]{polak-etal-2022-cuni}
Peter Pol{\'a}k, Ngoc-Quan Pham, Tuan~Nam Nguyen, Danni Liu, Carlos Mullov, Jan Niehues, Ond{\v{r}}ej Bojar, and Alexander Waibel. 2022.
\newblock \href {https://doi.org/10.18653/v1/2022.iwslt-1.24} {{CUNI}-{KIT} system for simultaneous speech translation task at {IWSLT} 2022}.
\newblock In \emph{Proceedings of the 19th International Conference on Spoken Language Translation (IWSLT 2022)}, pages 277--285, Dublin, Ireland (in-person and online).

\bibitem[{Post(2018)}]{post-2018-call}
Matt Post. 2018.
\newblock \href {https://www.aclweb.org/anthology/W18-6319} {{A Call for Clarity in Reporting {BLEU} Scores}}.
\newblock In \emph{Proceedings of the Third Conference on Machine Translation: Research Papers}, pages 186--191, Belgium, Brussels.

\bibitem[{Ren et~al.(2020)Ren, Liu, Tan, Zhang, Qin, Zhao, and Liu}]{ren-etal-2020-simulspeech}
Yi~Ren, Jinglin Liu, Xu~Tan, Chen Zhang, Tao Qin, Zhou Zhao, and Tie-Yan Liu. 2020.
\newblock \href {https://doi.org/10.18653/v1/2020.acl-main.350} {{S}imul{S}peech: End-to-end simultaneous speech to text translation}.
\newblock In \emph{Proceedings of the 58th Annual Meeting of the Association for Computational Linguistics}, pages 3787--3796, Online.

\bibitem[{Szegedy et~al.(2016)Szegedy, Vanhoucke, Ioffe, Shlens, and Wojna}]{szegedy2016rethinking}
Christian Szegedy, Vincent Vanhoucke, Sergey Ioffe, Jon Shlens, and Zbigniew Wojna. 2016.
\newblock {Rethinking the Inception Architecture for Computer Vision}.
\newblock In \emph{Proc. of 2016 IEEE CVPR}, pages 2818--2826, Las Vegas, Nevada, United States.

\bibitem[{Tang et~al.(2018)Tang, Sennrich, and Nivre}]{tang-etal-2018-analysis}
Gongbo Tang, Rico Sennrich, and Joakim Nivre. 2018.
\newblock \href {https://doi.org/10.18653/v1/W18-6304} {An analysis of attention mechanisms: The case of word sense disambiguation in neural machine translation}.
\newblock In \emph{Proceedings of the Third Conference on Machine Translation: Research Papers}, pages 26--35, Brussels, Belgium.

\bibitem[{Tang et~al.(2023)Tang, Sun, Inaguma, Chen, Dong, Ma, Tomasello, and Pino}]{tang-etal-2023-hybrid}
Yun Tang, Anna Sun, Hirofumi Inaguma, Xinyue Chen, Ning Dong, Xutai Ma, Paden Tomasello, and Juan Pino. 2023.
\newblock \href {https://doi.org/10.18653/v1/2023.acl-long.695} {Hybrid transducer and attention based encoder-decoder modeling for speech-to-text tasks}.
\newblock In \emph{Proceedings of the 61st Annual Meeting of the Association for Computational Linguistics (Volume 1: Long Papers)}, pages 12441--12455, Toronto, Canada.

\bibitem[{Tsiamas et~al.(2023)Tsiamas, Fonollosa, and Costa-juss{\`a}}]{tsiamas-etal-2023-segaugment}
Ioannis Tsiamas, Jos{\'e} Fonollosa, and Marta Costa-juss{\`a}. 2023.
\newblock \href {https://doi.org/10.18653/v1/2023.findings-emnlp.574} {{S}eg{A}ugment: Maximizing the utility of speech translation data with segmentation-based augmentations}.
\newblock In \emph{Findings of the Association for Computational Linguistics: EMNLP 2023}, pages 8569--8588, Singapore. Association for Computational Linguistics.

\bibitem[{Vaswani et~al.(2017)Vaswani, Shazeer, Parmar, Uszkoreit, Jones, Gomez, Kaiser, and Polosukhin}]{transformer}
Ashish Vaswani, Noam Shazeer, Niki Parmar, Jakob Uszkoreit, Llion Jones, Aidan~N Gomez, \L~ukasz Kaiser, and Illia Polosukhin. 2017.
\newblock \href {https://proceedings.neurips.cc/paper/2017/file/3f5ee243547dee91fbd053c1c4a845aa-Paper.pdf} {Attention is all you need}.
\newblock In \emph{Advances in Neural Information Processing Systems}, volume~30.

\bibitem[{Wang et~al.(2020)Wang, Tang, Ma, Wu, Okhonko, and Pino}]{wang2020fairseqs2t}
Changhan Wang, Yun Tang, Xutai Ma, Anne Wu, Dmytro Okhonko, and Juan Pino. 2020.
\newblock fairseq s2t: Fast speech-to-text modeling with fairseq.
\newblock In \emph{Proceedings of the 2020 Conference of the Asian Chapter of the Association for Computational Linguistics (AACL): System Demonstrations}.

\bibitem[{Weiss et~al.(2017)Weiss, Chorowski, Jaitly, Wu, and Chen}]{weiss2017sequence}
Ron~J. Weiss, Jan Chorowski, Navdeep Jaitly, Yonghui Wu, and Zhifeng Chen. 2017.
\newblock {Sequence-to-Sequence Models Can Directly Translate Foreign Speech}.
\newblock In \emph{Proceedings of Interspeech 2017}, pages 2625--2629, Stockholm, Sweden.

\bibitem[{Weller et~al.(2021)Weller, Sperber, Gollan, and Kluivers}]{weller-etal-2021-streaming}
Orion Weller, Matthias Sperber, Christian Gollan, and Joris Kluivers. 2021.
\newblock \href {https://doi.org/10.18653/v1/2021.eacl-main.216} {Streaming models for joint speech recognition and translation}.
\newblock In \emph{Proceedings of the 16th Conference of the European Chapter of the Association for Computational Linguistics: Main Volume}, pages 2533--2539, Online.

\bibitem[{Zeng et~al.(2021)Zeng, Li, and Liu}]{zeng-etal-2021-realtrans}
Xingshan Zeng, Liangyou Li, and Qun Liu. 2021.
\newblock \href {https://doi.org/10.18653/v1/2021.findings-acl.218} {{R}eal{T}ran{S}: End-to-end simultaneous speech translation with convolutional weighted-shrinking transformer}.
\newblock In \emph{Findings of the Association for Computational Linguistics: ACL-IJCNLP 2021}, pages 2461--2474, Online.

\bibitem[{Zeng et~al.(2022)Zeng, Li, Li, and Liu}]{zeng-etal-2022-end}
Xingshan Zeng, Pengfei Li, Liangyou Li, and Qun Liu. 2022.
\newblock \href {https://doi.org/10.18653/v1/2022.autosimtrans-1.5} {End-to-end simultaneous speech translation with pretraining and distillation: Huawei {N}oah{'}s system for {A}uto{S}im{T}ran{S} 2022}.
\newblock In \emph{Proceedings of the Third Workshop on Automatic Simultaneous Translation}, pages 25--33, Online.

\bibitem[{Zenkel et~al.(2019)Zenkel, Wuebker, and DeNero}]{zenkel2019adding}
Thomas Zenkel, Joern Wuebker, and John DeNero. 2019.
\newblock Adding interpretable attention to neural translation models improves word alignment.
\newblock \emph{arXiv preprint arXiv:1901.11359}.

\bibitem[{Zhang et~al.(2022)Zhang, He, Wu, and Wang}]{zhang-etal-2022-learning}
Ruiqing Zhang, Zhongjun He, Hua Wu, and Haifeng Wang. 2022.
\newblock \href {https://doi.org/10.18653/v1/2022.acl-long.542} {Learning adaptive segmentation policy for end-to-end simultaneous translation}.
\newblock In \emph{Proceedings of the 60th Annual Meeting of the Association for Computational Linguistics (Volume 1: Long Papers)}, pages 7862--7874, Dublin, Ireland.

\bibitem[{Zhang and Feng(2022)}]{zhang2022information}
Shaolei Zhang and Yang Feng. 2022.
\newblock \href {https://aclanthology.org/2022.emnlp-main.65} {Information-transport-based policy for simultaneous translation}.
\newblock In \emph{Proceedings of the 2022 Conference on Empirical Methods in Natural Language Processing}, pages 992--1013, Abu Dhabi, United Arab Emirates.

\bibitem[{Zhang and Feng(2023)}]{zhang-feng-2023-end}
Shaolei Zhang and Yang Feng. 2023.
\newblock \href {https://doi.org/10.18653/v1/2023.findings-acl.485} {End-to-end simultaneous speech translation with differentiable segmentation}.
\newblock In \emph{Findings of the Association for Computational Linguistics: ACL 2023}, pages 7659--7680, Toronto, Canada.

\end{thebibliography}

\clearpage

\appendix

\section{The choice of $n_{words}$}
\label{app:fixed-words}


\begin{table*}[!h]
\setlength{\tabcolsep}{1.5pt}
\small
    \centering
    \begin{tabular}{c||ccc|ccc|ccc|ccc||ccc}
    \specialrule{.1em}{.05em}{.05em}
        \multirow{3}{*}{history} & \multicolumn{3}{c|}{$f=2$} & \multicolumn{3}{c|}{$f=4$} & \multicolumn{3}{c|}{$f=6$} & \multicolumn{3}{c||}{$f=8$} & \multicolumn{3}{c}{\textbf{AVG}} \\
         \cline{2-16}
         & \multirow{2}{*}{BLEU} & \multicolumn{2}{c|}{StreamLAAL} &
         \multirow{2}{*}{BLEU} & \multicolumn{2}{c|}{StreamLAAL} &
         \multirow{2}{*}{BLEU} & \multicolumn{2}{c|}{StreamLAAL} & \multirow{2}{*}{BLEU} & \multicolumn{2}{c||}{StreamLAAL} & \multirow{2}{*}{BLEU} & \multicolumn{2}{c}{StreamLAAL}\\
         \cline{3-4} \cline{6-7} \cline{9-10} \cline{12-13} \cline{15-16}
         & & NCA & CA &  & NCA & CA &  & NCA & CA &  & NCA & CA &   & NCA & CA \\
         \hline
        $n_{words}=10$ & 21.7 & 2.20 & 3.66 & 23.8 & 2.45 & 3.82 & 24.7 & 2.69 & 4.03 & 24.8 & 2.98 & 4.27 & 23.8 & 2.58 & 3.95 \\
        $n_{words}=20$ & 21.2 & 1.64 & 3.09 & 23.5 & 1.96 & 3.28 & 24.4 & 2.37 & 4.05 & 24.9 & 2.71 & 4.37 & 23.5 & 2.17 & 3.70 \\
        $n_{words}=30$ & 21.1 & 1.52 & 3.24 & 23.2 & 1.98 & 3.97 & 24.3 & 2.18 & 3.84 & 24.6 & 2.62 & 4.54 & 23.3 & 2.08 & 3.90 \\
        $n_{words}=40$ & 20.9 & 1.58 & 3.83 & 23.3 & 1.87 & 3.93 & 24.4 & 2.21 & 4.27 & 25.0 & 2.51 & 4.66 & 23.4 & 2.04 & 4.17 \\
        \specialrule{.1em}{.05em}{.05em}
    \end{tabular}
    \caption{StreamAttFW results on MuST-C en-de dev set.}
    \label{tab:nwords-streamatt}
\end{table*}

Looking at the results of Table \ref{tab:nwords-streamatt}, it emerges that although the solution with $n_{words}=10$ achieves the highest average BLEU score (23.8), it is the slowest in terms of latency. This counter-intuitive behavior can be explained by the fact that with a reduced history context, the model tends to wait longer before generating a partial translation, thereby improving output quality but impacting latency.  Conversely, with $n_{words}=30$ and $n_{words}=40$, we achieve lower latency scores (with $n_{words}=40$ showing a slightly lower non-computational-aware StreamLAAL but a 
higher computational-aware StreamLAAL), at 
a slight detriment 
in translation quality. In this case, the behavior can be attributed to the increased history context, which makes the model more confident in its hypothesis, resulting in earlier translation emission compared to the case with a reduced history, albeit with a small quality degradation. As a result, we select the solution with $n_{words}=20$ that represents the better trade-off between quality and latency since it obtains the 
best quality-latency ratio\footnote{$\frac{\text{BLEU}}{\text{StreamLAAL}_{\text{NCA}}+\text{StreamLAAL}_{\text{CA}}}$} of about 4.0 against 3.6 ($n_{words}=10$), 3.9 ($n_{words}=30$), and 3.8 ($n_{words}=40$).

\section{NLLB 3.3B performance on MuST-C}
\label{app:nllb}
See Table \ref{tab:nllb}.

\begin{table}[!h]
\small
    \centering
    \setlength{\tabcolsep}{4pt}
    \begin{tabular}{ccccccccc}
    \specialrule{.1em}{.05em}{.05em} 
        de & es & fr & it & nl & pt & ro & ru & Avg \\
        \hline
        33.1 & 38.5 & 46.5 & 34.4 & 37.7 & 40.4 & 32.8 & 23.5 & 35.9 \\
    \specialrule{.1em}{.05em}{.05em} 
    \end{tabular}
    \caption{BLEU results of the NLLB 3.3B model on all the language pairs of MuST-C v1.0 tst-COMMON.}
    \label{tab:nllb}
\end{table}

\section{Model and Training Settings}
\label{sec:train_setup}

The Conformer-based model is made of 12 Conformer encoder layers \citep{gulati20_interspeech} and 6 Transformer \citep{transformer} decoder layers with a total of $\sim$115M parameters.
Each encoder/decoder layer has 8 attention heads, 
512 as embedding size and 2,048 hidden neurons in the feed-forward layers. 
We set dropout at 0.1 for feed-forward, attention, and convolution layers. Also, in the convolution layer, we set 31 as the kernel size for the point- and depth-wise convolutions. 
The vocabularies are based on unigram SentencePiece models \citep{kudo-richardson-2018-sentencepiece} with dimensions of 8,000 for the target side and 5,000 for the source side (en).
We optimize with Adam \citep{DBLP:journals/corr/KingmaB14} by using the label-smoothed cross-entropy loss with 0.1 as the smoothing factor \citep{szegedy2016rethinking}. We employ Connectionist Temporal Classification -- or CTC -- \citep{Graves2006ConnectionistTC} as an auxiliary loss to avoid pre-training \citep{gaido-etal-2022-efficient} and also to compress the input audio, reducing RAM consumption and speeding up inference \citep{gaido-etal-2021-ctc}.
Utterance-level Cepstral Mean and Variance Normalization (CMVN) and SpecAugment \citep{Park2019} are applied during training. 
The learning rate is set to $5\cdot10^{-3}$ with Noam scheduler \citep{transformer} and warm-up steps of 25k. 
We stop the training after 15 epochs without loss decrease on the dev set and average 7 checkpoints around the best (best, three preceding, and three succeeding). 
Trainings are performed on 4 NVIDIA A40 GPUs with 40GB RAM. We set 40k as the maximum number of tokens per mini-batch, 2 as
update frequency, and 100,000 as maximum updates ($\sim$23 hours).

\section{Streaming Results per Language}
\label{app:steam-per-lang}

See Table \ref{tab:stream-res-lang}.

\begin{table*}[!ht]
\setlength{\tabcolsep}{1.5pt}
\small
    \centering
    \begin{tabular}{l||ccc|ccc|ccc|ccc||ccc}
    \specialrule{.1em}{.05em}{.05em}
    \multicolumn{16}{c}{\textbf{en-de}} \\
    \specialrule{.1em}{.05em}{.05em}
        \multirow{3}{*}{Strategy} & \multicolumn{3}{c|}{$f=2$} & \multicolumn{3}{c|}{$f=4$} & \multicolumn{3}{c|}{$f=6$} & \multicolumn{3}{c||}{$f=8$} & \multicolumn{3}{c}{\textbf{AVG}} \\
         \cline{2-16}
         & \multirow{2}{*}{BLEU} & \multicolumn{2}{c|}{StreamLAAL} &
         \multirow{2}{*}{BLEU} & \multicolumn{2}{c|}{StreamLAAL} &
         \multirow{2}{*}{BLEU} & \multicolumn{2}{c|}{StreamLAAL} & \multirow{2}{*}{BLEU} & \multicolumn{2}{c||}{StreamLAAL} & \multirow{2}{*}{BLEU} & \multicolumn{2}{c}{StreamLAAL}\\
         \cline{3-4} \cline{6-7} \cline{9-10} \cline{12-13} \cline{15-16}
         & & NCA & CA &  & NCA & CA &  & NCA & CA &  & NCA & CA &   & NCA & CA \\
         \hline
        Baseline & 16.8 & 3.01 & 4.97 & 17.8 & 3.09 & 5.03 & 18.3 & 3.34 & 5.40 & 18.3 & 3.82 & 5.81 & 17.8 & 3.32 & 5.30 \\
        StreamAttFW & 21.1 & 1.50 & 3.03 & 23.0 & 1.81 & 3.23 & 23.8 & 2.06 & 3.45 & 24.6 & 2.36 & 3.67 & 23.1 & 1.93 & 3.35 \\
        StreamAttP & 20.7 & 1.53 & 3.46 & 22.5 & 1.87 & 3.74 & 23.5 & 2.13 & 3.97 & 23.9 & 2.43 & 4.18 & 22.7 & 1.99 & 3.84 \\
    \specialrule{.1em}{.05em}{.05em}
        \multicolumn{16}{c}{\textbf{en-es}} \\
    \specialrule{.1em}{.05em}{.05em}
        \multirow{3}{*}{Strategy} & \multicolumn{3}{c|}{$f=2$} & \multicolumn{3}{c|}{$f=4$} & \multicolumn{3}{c|}{$f=6$} & \multicolumn{3}{c||}{$f=8$} & \multicolumn{3}{c}{\textbf{AVG}} \\
         \cline{2-16}
         & \multirow{2}{*}{BLEU} & \multicolumn{2}{c|}{StreamLAAL} &
         \multirow{2}{*}{BLEU} & \multicolumn{2}{c|}{StreamLAAL} &
         \multirow{2}{*}{BLEU} & \multicolumn{2}{c|}{StreamLAAL} & \multirow{2}{*}{BLEU} & \multicolumn{2}{c||}{StreamLAAL} & \multirow{2}{*}{BLEU} & \multicolumn{2}{c}{StreamLAAL}\\
         \cline{3-4} \cline{6-7} \cline{9-10} \cline{12-13} \cline{15-16}
         & & NCA & CA &  & NCA & CA &  & NCA & CA &  & NCA & CA &   & NCA & CA \\
         \hline
        Baseline & 20.7 & 2.19 & 3.57 & 21.4 & 2.36 & 3.83 & 22.0 & 2.56 & 3.95 & 22.4 & 2.92 & 4.37 & 21.6 & 2.51 & 3.93 \\
        StreamAttFW & 24.4 & 1.27 & 2.39 & 26.2 & 1.57 & 2.70 & 27.3 & 1.85 & 2.92 & 27.8 & 2.15 & 3.24 & 26.4 & 1.71 & 2.81 \\
        StreamAttP & 25.2 & 1.49 & 3.28 & 26.7 & 1.80 & 3.77 & 27.0 & 2.01 & 3.90 & 27.6 & 2.39 & 4.36 & 26.6 & 1.92 & 3.83 \\
    \specialrule{.1em}{.05em}{.05em}
        \multicolumn{16}{c}{\textbf{en-fr}} \\
    \specialrule{.1em}{.05em}{.05em}
        \multirow{3}{*}{Strategy} & \multicolumn{3}{c|}{$f=2$} & \multicolumn{3}{c|}{$f=4$} & \multicolumn{3}{c|}{$f=6$} & \multicolumn{3}{c||}{$f=8$} & \multicolumn{3}{c}{\textbf{AVG}} \\
         \cline{2-16}
         & \multirow{2}{*}{BLEU} & \multicolumn{2}{c|}{StreamLAAL} &
         \multirow{2}{*}{BLEU} & \multicolumn{2}{c|}{StreamLAAL} &
         \multirow{2}{*}{BLEU} & \multicolumn{2}{c|}{StreamLAAL} & \multirow{2}{*}{BLEU} & \multicolumn{2}{c||}{StreamLAAL} & \multirow{2}{*}{BLEU} & \multicolumn{2}{c}{StreamLAAL}\\
         \cline{3-4} \cline{6-7} \cline{9-10} \cline{12-13} \cline{15-16}
         & & NCA & CA &  & NCA & CA &  & NCA & CA &  & NCA & CA &   & NCA & CA \\
         \hline
        Baseline & 26.2 & 2.36 & 3.90 & 27.7 & 2.64 & 4.23 & 28.6 & 2.84 & 4.36 & 28.8 & 3.20 & 4.89 & 27.8 & 2.76 & 4.35 \\
        StreamAttFW & 30.1 & 1.46 & 2.76 & 33.0 & 1.68 & 2.95 & 34.3 & 2.02 & 3.25 & 35.0 & 2.26 & 3.47 & 33.1 & 1.86 & 3.12 \\
        StreamAttP & 31.6 & 1.47 & 3.18 & 33.6 & 1.74 & 3.51 & 34.5 & 2.11 & 3.85 & 35.1 & 2.41 & 4.12 & 33.7 & 1.93 & 3.67 \\
    \specialrule{.1em}{.05em}{.05em}
        \multicolumn{16}{c}{\textbf{en-it}} \\
    \specialrule{.1em}{.05em}{.05em}
        \multirow{3}{*}{Strategy} & \multicolumn{3}{c|}{$f=2$} & \multicolumn{3}{c|}{$f=4$} & \multicolumn{3}{c|}{$f=6$} & \multicolumn{3}{c||}{$f=8$} & \multicolumn{3}{c}{\textbf{AVG}} \\
         \cline{2-16}
         & \multirow{2}{*}{BLEU} & \multicolumn{2}{c|}{StreamLAAL} &
         \multirow{2}{*}{BLEU} & \multicolumn{2}{c|}{StreamLAAL} &
         \multirow{2}{*}{BLEU} & \multicolumn{2}{c|}{StreamLAAL} & \multirow{2}{*}{BLEU} & \multicolumn{2}{c||}{StreamLAAL} & \multirow{2}{*}{BLEU} & \multicolumn{2}{c}{StreamLAAL}\\
         \cline{3-4} \cline{6-7} \cline{9-10} \cline{12-13} \cline{15-16}
         & & NCA & CA &  & NCA & CA &  & NCA & CA &  & NCA & CA &   & NCA & CA \\
         \hline
        Baseline & 16.9 & 2.21 & 3.74 & 17.7 & 2.55 & 4.03 & 18.1 & 3.14 & 4.68 & 18.0 & 3.41 & 4.93 & 17.7 & 2.83 & 4.35 \\
        StreamAttFW & 20.7 & 1.30 & 2.51 & 22.6 & 1.58 & 2.76 & 23.4 & 1.90 & 3.03 & 23.8 & 2.22 & 3.30 & 22.6 & 1.75 & 2.90 \\
        StreamAttP & 21.0 & 1.51 & 3.25 & 21.7 & 1.71 & 3.49 & 23.0 & 2.09 & 3.75 & 23.2 & 2.39 & 3.99 & 22.2 & 1.93 & 3.62 \\
    \specialrule{.1em}{.05em}{.05em}
        \multicolumn{16}{c}{\textbf{en-nl}} \\
    \specialrule{.1em}{.05em}{.05em}
        \multirow{3}{*}{Strategy} & \multicolumn{3}{c|}{$f=2$} & \multicolumn{3}{c|}{$f=4$} & \multicolumn{3}{c|}{$f=6$} & \multicolumn{3}{c||}{$f=8$} & \multicolumn{3}{c}{\textbf{AVG}} \\
         \cline{2-16}
         & \multirow{2}{*}{BLEU} & \multicolumn{2}{c|}{StreamLAAL} &
         \multirow{2}{*}{BLEU} & \multicolumn{2}{c|}{StreamLAAL} &
         \multirow{2}{*}{BLEU} & \multicolumn{2}{c|}{StreamLAAL} & \multirow{2}{*}{BLEU} & \multicolumn{2}{c||}{StreamLAAL} & \multirow{2}{*}{BLEU} & \multicolumn{2}{c}{StreamLAAL}\\
         \cline{3-4} \cline{6-7} \cline{9-10} \cline{12-13} \cline{15-16}
         & & NCA & CA &  & NCA & CA &  & NCA & CA &  & NCA & CA &   & NCA & CA \\
         \hline
        Baseline & 20.0 & 2.69 & 4.35 & 20.8 & 2.91 & 4.62 & 21.1 & 2.94 & 4.69 & 21.2 & 3.39 & 5.22 & 20.8 & 2.98 & 4.72 \\
        StreamAttFW & 24.0 & 1.47 & 2.79 & 26.1 & 1.74 & 3.02 & 26.9 & 2.00 & 3.33 & 27.4 & 2.28 & 3.59 & 26.1 & 1.87 & 3.18 \\
        StreamAttP & 23.2 & 2.39 & 3.99 & 25.7 & 1.95 & 4.07 & 26.5 & 2.14 & 4.17 & 27.1 & 2.50 & 4.50 & 25.6 & 2.25 & 4.18 \\
    \specialrule{.1em}{.05em}{.05em}
        \multicolumn{16}{c}{\textbf{en-pt}} \\
    \specialrule{.1em}{.05em}{.05em}
        \multirow{3}{*}{Strategy} & \multicolumn{3}{c|}{$f=2$} & \multicolumn{3}{c|}{$f=4$} & \multicolumn{3}{c|}{$f=6$} & \multicolumn{3}{c||}{$f=8$} & \multicolumn{3}{c}{\textbf{AVG}} \\
         \cline{2-16}
         & \multirow{2}{*}{BLEU} & \multicolumn{2}{c|}{StreamLAAL} &
         \multirow{2}{*}{BLEU} & \multicolumn{2}{c|}{StreamLAAL} &
         \multirow{2}{*}{BLEU} & \multicolumn{2}{c|}{StreamLAAL} & \multirow{2}{*}{BLEU} & \multicolumn{2}{c||}{StreamLAAL} & \multirow{2}{*}{BLEU} & \multicolumn{2}{c}{StreamLAAL}\\
         \cline{3-4} \cline{6-7} \cline{9-10} \cline{12-13} \cline{15-16}
         & & NCA & CA &  & NCA & CA &  & NCA & CA &  & NCA & CA &   & NCA & CA \\
         \hline
        Baseline & 21.0 & 2.56 & 4.11 & 21.6 & 2.89 & 4.44 & 22.1 & 2.89 & 4.54 & 22.1 & 3.77 & 5.36 & 21.7 & 3.03 & 4.61 \\
        StreamAttFW & 25.5 & 1.43 & 2.64 & 27.6 & 1.74 & 2.90 & 28.3 & 2.01 & 3.22 & 28.9 & 2.26 & 3.38 & 27.6 & 1.86 & 3.04 \\
        StreamAttP & 26.4 & 1.72 & 3.87 & 27.8 & 1.92 & 4.20 & 28.2 & 2.20 & 4.28 & 28.7 & 2.56 & 4.72 & 27.8 & 2.10 & 4.28 \\
    \specialrule{.1em}{.05em}{.05em}
        \multicolumn{16}{c}{\textbf{en-ro}} \\
    \specialrule{.1em}{.05em}{.05em}
        \multirow{3}{*}{Strategy} & \multicolumn{3}{c|}{$f=2$} & \multicolumn{3}{c|}{$f=4$} & \multicolumn{3}{c|}{$f=6$} & \multicolumn{3}{c||}{$f=8$} & \multicolumn{3}{c}{\textbf{AVG}} \\
         \cline{2-16}
         & \multirow{2}{*}{BLEU} & \multicolumn{2}{c|}{StreamLAAL} &
         \multirow{2}{*}{BLEU} & \multicolumn{2}{c|}{StreamLAAL} &
         \multirow{2}{*}{BLEU} & \multicolumn{2}{c|}{StreamLAAL} & \multirow{2}{*}{BLEU} & \multicolumn{2}{c||}{StreamLAAL} & \multirow{2}{*}{BLEU} & \multicolumn{2}{c}{StreamLAAL}\\
         \cline{3-4} \cline{6-7} \cline{9-10} \cline{12-13} \cline{15-16}
         & & NCA & CA &  & NCA & CA &  & NCA & CA &  & NCA & CA &   & NCA & CA \\
         \hline
        Baseline & 16.1 & 2.20 & 3.64 & 16.7 & 2.66 & 4.16 & 17.2 & 2.80 & 4.29 & 17.2 & 3.30 & 4.84 & 16.8 & 2.74 & 4.23 \\
        StreamAttFW & 19.6 & 1.29 & 2.53 & 21.0 & 1.66 & 2.86 & 21.7 & 1.94 & 3.14 & 22.2 & 2.25 & 3.42 & 21.1 & 1.79 & 2.99 \\
        StreamAttP & 20.1 & 1.57 & 3.40 & 21.6 & 1.88 & 3.62 & 22.1 & 2.30 & 4.21 & 22.5 & 2.52 & 4.42 & 21.6 & 2.07 & 3.91 \\
    \specialrule{.1em}{.05em}{.05em}
        \multicolumn{16}{c}{\textbf{en-ru}} \\
    \specialrule{.1em}{.05em}{.05em}
        \multirow{3}{*}{Strategy} & \multicolumn{3}{c|}{$f=2$} & \multicolumn{3}{c|}{$f=4$} & \multicolumn{3}{c|}{$f=6$} & \multicolumn{3}{c||}{$f=8$} & \multicolumn{3}{c}{\textbf{AVG}} \\
         \cline{2-16}
         & \multirow{2}{*}{BLEU} & \multicolumn{2}{c|}{StreamLAAL} &
         \multirow{2}{*}{BLEU} & \multicolumn{2}{c|}{StreamLAAL} &
         \multirow{2}{*}{BLEU} & \multicolumn{2}{c|}{StreamLAAL} & \multirow{2}{*}{BLEU} & \multicolumn{2}{c||}{StreamLAAL} & \multirow{2}{*}{BLEU} & \multicolumn{2}{c}{StreamLAAL}\\
         \cline{3-4} \cline{6-7} \cline{9-10} \cline{12-13} \cline{15-16}
         & & NCA & CA &  & NCA & CA &  & NCA & CA &  & NCA & CA &   & NCA & CA \\
         \hline
        Baseline & 10.3 & 3.94 & 7.01 & 10.6 & 4.27 & 7.73 & 10.7 & 4.05 & 7.47 & 10.8 & 4.94 & 8.68 & 10.6 & 4.30 & 7.72 \\
        StreamAttFW & 13.2 & 1.65 & 4.07 & 14.7 & 1.88 & 3.90 & 15.3 & 2.25 & 4.35 & 15.4 & 2.59 & 4.85 & 14.7 & 2.09 & 4.29 \\
        StreamAttP & 13.6 & 1.56 & 3.85 & 14.8 & 1.81 & 4.10 & 15.1 & 2.22 & 4.52 & 15.3 & 2.55 & 4.87 & 14.7 & 2.04 & 4.34 \\
    \specialrule{.1em}{.05em}{.05em}
    \end{tabular}
    \caption{Quality (BLEU$\uparrow$), non-computationally and computationally aware (NCA/CA) latency (StreamLAAL$\downarrow$) results on MuST-C v1.0 tst-COMMON for all the 8 languages.}
    \label{tab:stream-res-lang}
\end{table*}



\end{document}